%% file: flavon_final.tex
\documentclass[preprintnumbers,11pt,superscriptaddress,endnote,nofootinbib,aps,prd,floatfix,tightenlines,a4paper]{revtex4-1}
\usepackage{amssymb,amsmath,multirow,graphicx,tabularx,epsfig}
\usepackage{multirow}
\usepackage{color}
\usepackage{slashed}
\usepackage{verbatim}
\usepackage[normalem]{ulem}
\usepackage{subfigure}
\usepackage{hyperref}
\usepackage{pgf}
\usepackage{booktabs}
\newcommand{\ra}[1]{\renewcommand{\arraystretch}{#1}}
\newcommand*{\myalign}[2]{\multicolumn{1}{#1}{#2}}
\usepackage{tikz}
\usetikzlibrary{arrows,automata}
\usetikzlibrary{positioning}
\usetikzlibrary{arrows,decorations.markings,decorations.pathmorphing}
\graphicspath{{figs/}}

\input{declare}

\usepackage{array}
\newcolumntype{P}[1]{>{\centering\arraybackslash}p{#1}}

\begin{document}

\title{Hunting the Flavon}

\author{Martin Bauer}
\affiliation{Institut f\"ur Theoretische Physik, Universit\"at Heidelberg, Germany}

\author{Torben Schell}
\affiliation{Institut f\"ur Theoretische Physik, Universit\"at Heidelberg, Germany}

\author{Tilman Plehn}
\affiliation{Institut f\"ur Theoretische Physik, Universit\"at Heidelberg, Germany}

\date{\today}

\begin{abstract} 
The next generation of experiments in particle physics will for the
first time systematically test flavor physics models based on flavon
fields.  Starting from the current quark-flavor constrains on such
models we show how the new generation of lepton flavor experiments
will dominate indirect searches in the coming decades.  A future 100
TeV hadron collider will then be the first experiment to probe flavons
as propagating degrees of freedom. Our estimate of the collider reach
relies on a proper treatment of backgrounds and detector effects.
Complementary searches for indirect effects in lepton flavor
experiments and propagating degrees of freedom at colliders are very
limited at the LHC, but will be a new feature at a 100 TeV hadron
collider.
\end{abstract}

\maketitle
\tableofcontents

\clearpage

\section{Introduction}
\label{sec:intro}

The structure of the quark and lepton flavor sectors is one of the
biggest mysteries of particle physics. Various extensions of the
Standard Model (SM) address the flavor structure for example through
abelian flavor symmetries~\cite{froggatt_nielsen,flavor_symm},
loop-suppressed couplings to the Higgs~\cite{higgs_coup}, partial
compositeness~\cite{partial_comp}, or wave-function
localization~\cite{wf_local}. All of these mechanisms introduce
flavor-violating couplings and new, heavy degrees of freedom.  For
instance, partial compositeness or warped extra dimensions predict
vector-like heavy quarks and colored spin-one resonances with large
cross sections, with features which are unfortunately not unique to
flavor models. Similar structures appear in alternative models.
However, experimental results drive the underlying mass scales
into regions which are not accessible by the LHC.  This is the reason
why theories for quark and lepton flavor physics usually rely on an
effective field theory description, neglecting the effects of
actually new particles.\bigskip

We propose search strategies for the dynamic agent of flavor symmetry
breaking~\cite{flavon_pheno}, the flavon, at a future hadron
collider. Using a minimal Froggatt-Nielsen setup we only allow for
couplings directly related to the generation of the flavor
hierarchies. This means that a future discovery can directly probe the
underlying mechanism of flavor symmetry breaking. In general, the
dimensionless Yukawa couplings do not favor any underlying mass scale;
a low flavor breaking scale appears if we link the flavor breaking and
the electroweak scales~\cite{martin_marcela} or if dark matter
interactions are mediated by flavon
exchange~\cite{Calibbi:2015sfa}. In this paper we deliberately remain
agnostic about the ultraviolet completion and discuss the accessible
parameter space independent of model building aspects\footnote{This
  includes the obvious application of flavon models to the observed
  750~GeV excess which we cannot be bothered to work out
  (yet).}.\bigskip

We start by reviewing the most stringent flavor bounds, including
projections of current and future experiments testing the quark and
lepton sectors. In recent years, significant progress has been made in
testing the quark flavor structure at LHC, Belle, and BaBar. Future
searches will only slightly increase their sensitivity. On the other
hand, searches for lepton flavor effects~\cite{lepton_earlier} are
entering a golden era with MEG~II, Mu3e, DeeMe, COMET, and Mu2e. They
should improve existing limits by orders of magnitude. In our setup we
see how they will probe parameter regions far beyond the reach of
quark flavor physics.

Next, we discuss the discovery reach of the LHC and of a 100~TeV
hadron collider~\cite{nimatron}. We find that a 100~TeV hadron
collider will for the first time allow us to probe a sizeable part of
the flavon parameter space, \textsl{i.e.} giving us access to the
actual dynamic degrees of freedom in the flavor sector rather than
constraining its symmetry structure based on effective field theory.
This way, flavon searches add a qualitatively new aspect to the case
of a future proton-proton collider, including WIMP dark matter
searches~\cite{nimatron_wimp}, Higgs precision
measurements~\cite{nimatron_higgs}, searches for new heavy
particles~\cite{nimatron_heavy}, and testing mechanisms of
baryogenesis~\cite{nimatron_baryo}.

\section{Flavon model}
\label{sec:model}

In the simplest flavon setup the Higgs and all Standard Model fermions,
except for the top, carry charges under a global $U(1)$ or a discrete
subgroup. The top Yukawa coupling is then the only allowed
renormalizable Yukawa coupling. Introducing a complex scalar field $S$
with flavor charge $a_S=1$ we write
\begin{align}
- \lag_\text{Yukawa}
&= y^d_{ij} \left(\frac{S}{\Lambda}\right)^{n^d_{ij}}\, \overline{Q}_i\, H \, d_{R_j} 
 + y^u_{ij} \left(\frac{S}{\Lambda}\right)^{n^u_{ij}}\, \overline{Q}_i\, \widetilde H \, u_{R_j} \notag \\
&+ y^\ell_{ij} \left(\frac{S}{\Lambda}\right)^{n^\ell_{ij}}\, \overline{L}_i\, H \, \ell_{R_j} 
 + y^\nu_{ij} \left(\frac{S}{\Lambda}\right)^{n^\nu_{ij}}\, \overline{L}_i\, \widetilde H \, \nu_{R_j}
 + \text{h.c.} 
\label{eq:lag1}
\end{align}
The indices $i,j=1,2,3$ link the the fundamental Yukawa couplings
$y_{ij}$ with corresponding powers of $S/\Lambda$. The last
term assumes the presence of right-handed neutrinos. The field $S$
develops a VEV through a potential
\begin{align}
- \lag_\text{potential}
= - \mu_S^2\, S^* S 
  + \lambda_S \, (S^* S)^2 
  + b\, (S^2 + S^{* 2} ) 
  + \lambda_{HS}  (S^* S)  (H^\dagger H)+V(H) \; .
\label{eq:potential}
\end{align}
For now we neglect the portal interaction, $\lambda_{HS}=0$. In its
presence, Higgs--flavon mixing~\cite{higgs_flavon} and deviations of
the Higgs couplings become an alternative strategy to search for the
flavon.  Under the assumption $\lambda_{HS}=0$ the physical flavon
fields is defined by excitations around the VEV,
\begin{align}
S(x)=\frac{f + s(x) +i\, a(x)}{\sqrt{2}} \; .
\label{eq:flavonfield}
\end{align}
The masses of the scalar and pseudo-scalar components are given by 
\begin{align}
m_s = \mu_S\,= \sqrt{\lambda_S} f 
\qqquad \text{and} \qqquad
m_a= \sqrt{2 b} \; .
\label{eq:masses}
\end{align}
This means that the mass of the pseudo-scalar `pion' of flavor
breaking remains a free parameter. It if stays below the flavor scale
we can assume the mass hierarchy
\begin{align}
m_a < m_s \approx f < \Lambda \; .
\end{align}
The pseudoscalar component of the flavon is most likely the first
resonance we would encounter in a search for a mechanism behind the
flavor structure of the Standard Model. In an abuse of notation, we
will therefore refer to it as the \textsl{pseudoscalar
  flavon}.\bigskip

The ratio
$\eps$ of the VEV and the ultraviolet mass scale
$\Lambda$ describes the entire flavor structure of the Standard Model,
\begin{align}
\eps 
=\frac{f}{\sqrt{2} \Lambda} 
= \frac{1}{\Lambda} \; \sqrt{\frac{\mu_S^2}{2 \lambda_S}}
\qqquad \text{with} \qqquad 
v < f < \Lambda \; . 
\label{eq:def_eps} 
\end{align}
For our numerical analysis we assume that $\eps$ is identified with
the Cabibbo angle
\begin{align}
\eps = (V_\text{CKM})_{12} \approx 0.23 \; .
\end{align}
The fundamental Yukawa matrices are assumed to be anarchic and of
order one
\begin{align}
|y^{u,d,\ell}| \approx \begin{pmatrix} 
1 &\,\, 1&\,\, 1\\
1 &\,\, 1&\,\, 1\\
1 &\,\, 1&\,\, 1
\end{pmatrix}\; .
\end{align}
Following the Lagrangian given in Eq.\eqref{eq:lag1} the numbers of insertions
$n_{ij}$ generate the effective Yukawa couplings
\begin{align}
- \lag_\text{Yukawa}= 
  Y^d_{ij}\; \overline{Q}_i\, H \, d_{R_j} 
+ Y^u_{ij}\; \overline{Q}_i\, \widetilde H \, u_{R_j} 
+ Y^\ell_{ij}\; \overline{L}_i\, H \, \ell_{R_j} 
+ Y^\nu_{ij}\; \overline{L}_i\, \widetilde H \, \nu_{R_j} + \text{h.c.} \; ,
\label{eq:lag2}
\end{align}
with $Y_{ij} = y_{ij} \,\eps^{n_{ij}}$.\bigskip

\subsubsection*{Flavon couplings}

The exponents $n_{ij}$ of the ratio $S/\Lambda$ defined in
Eq.\eqref{eq:lag1} can be expressed in terms of the flavor charges of
the fermions and Higgs bosons. For the quarks they read
\begin{align}
n_{ij}^d &= a_{Q_i}-a_{d_j} - a_H \notag \\
n_{ij}^u &= a_{Q_i}-a_{u_j} + a_H \; ,
\end{align}
where $a_{u_j} = a_{u,c,t}$ and $a_{d_j} = a_{d,s,b}$ denote the
flavor charges of the three generations of quark singlets, $a_{Q_i}$
are the flavor charges of the three generations of quark doublets, and
$a_H$ is the flavor charge of the Higgs. To obtain the correct quark
masses in our benchmark scenario we set $a_S=+1$, $a_H=0$, and
\begin{align}\label{charges}
  \begin{pmatrix}
    a_{Q_1} & a_{Q_2} & a_{Q_3} \\
    a_{u} & a_{c} & a_{t}\\
    a_{d} & a_{s} & a_{b}
  \end{pmatrix}
= \begin{pmatrix}
     3 &  2 & 0 \\
    -5 & -2 & 0 \\
    -4 & -3 & -3
  \end{pmatrix}\,.
\end{align}
Combined with order-one Yukawa couplings, as spelled out in the
Appendix, this gives the quark masses
\begin{align}
m_t\approx \frac{v}{\sqrt{2}} \qquad 
\frac{m_b}{m_t} \approx \eps^3 \qquad 
\frac{m_c}{m_t} \approx \eps^4 \qquad  
\frac{m_s}{m_t} \approx \eps^5 \qquad 
\frac{m_d}{m_t} \approx \eps^7 \qquad 
\frac{m_u}{m_t} \approx \eps^8 \; ,
\label{eq:qmass}
\end{align}
and the CKM matrix becomes
\begin{align}
V_\text{CKM}
\approx \begin{pmatrix}
 1 & \eps & \eps^3 \\
 \eps & 1 & \eps^2 \\
 \eps^3 & \eps^2 & 1 \\
\end{pmatrix} \; ,
\label{eq:ckmcond2}
\end{align}
The flavon couplings to fermions in the mass eigenbasis are linked to
the Yukawa couplings,
\begin{align}
g_{af_{iL}f_{jR}}^{u} \equiv
g_{aij}^u = \frac{1}{f} \;
\begin{pmatrix} 
    8 m_u              &  \eps m_c  &  \epsilon^3 m_t \\ 
     \eps^3 m_c & 4 m_c              &  \epsilon^2 m_t \\
     \eps^5 m_t &  \eps^2 m_t   & 0
\end{pmatrix}
\qqquad 
g_{aij}^{d} = \frac{1}{f} \;
\begin{pmatrix} 
    7 m_d            &  \eps m_s &  \eps^3 m_b \\ 
     \eps m_s & 5 m_s              &  \eps^2 m_b \\
     \eps m_b &  \eps^2 m_b   & 3 m_b
\end{pmatrix} \; .
\label{eq:qcoup}
\end{align}
where in the off-diagonal terms we neglect order-one factors. The fact
that the flavon does not couple to top quarks reflects our assumption
that the corresponding term in the Lagrangian starts at $\eps^0$, \textsl{i.e.}
without any suppression $f/\Lambda$.\bigskip

In the lepton sector the analogous exponents in Eq.\eqref{eq:lag1} are 
given by
\begin{align}
n_{ij}^\ell &= a_{L_i}-a_{\ell_j} - a_H \notag \\
n_{ij}^\nu &= a_{L_i}-a_{\nu_j} + a_H\; ,
\end{align}
in terms of the ten flavor charges.  As in the quark sector, we choose
the charges to reproduce the lepton masses and mixing patterns,
\begin{align}\label{lepcharges}
 \begin{pmatrix}
    a_{L_1} & a_{L_2} & a_{L_3} \\
   a_{\nu_e} & a_{\nu_\mu} & a_{\nu_\tau}\\
      a_{e} & a_{\mu} & a_{\tau} 
  \end{pmatrix}
= \begin{pmatrix}
    1 & 0 & 0 \\
    -24 & -21 & -20 \\
    -8 & -5 & -3
  \end{pmatrix} \; .
\end{align}
The neutrino charges can be smaller if a Majorana mass term
exists. One attractive way to implement it is to assume a flavor
charge of $\nu_R = 1/2$, such that
\begin{align}
\lag_\text{Majorana} = M_\nu\, \nu_R \nu_R\,, 
\end{align}
with $M_\nu = f$. This gives us the lepton mass ratios
\begin{align}
\frac{m_\tau}{m_t} \approx \eps^3 \qquad 
\frac{m_\mu}{m_t} \approx \eps^5 \qquad  
\frac{m_e}{m_t} \approx \eps^9 \qquad 
\frac{m_{\nu_1}}{m_t} \approx \eps^{25} \qquad 
\frac{m_{\nu_2}}{m_t} \approx \eps^{21} \qquad 
\frac{m_{\nu_3}}{m_t} \approx \eps^{20} \; ,
\label{eq:qmass}
\end{align}
and the leptonic mixing matrix 
\begin{align}
U_\text{PMNS}
\approx \begin{pmatrix}
 1 & \epsilon & \epsilon \\
 \epsilon & 1 & 1 \\
 \epsilon & 1 & 1 \\
\end{pmatrix} \; .
\end{align}
Again, the flavon couplings are related to the Yukawa couplings,
modulo order-one corrections in the off-diagonal terms of
\begin{align}
g_{af_{iL}f_{jR}}^{\ell} \equiv
g_{aij}^{\ell} = \frac{1}{f} \;
\begin{pmatrix}
    9 m_e                 &  \eps m_\mu    &  \eps m_\tau\\
     \eps^3 m_\mu^3    & 5 m_\mu             &  \eps^2 m_\tau\\
     \eps^5 m_\tau &  \eps^2 m_\tau & 3 m_\tau
  \end{pmatrix}
\label{eq:lcoup}
\end{align}
In all cases the corresponding scalar couplings to fermions, except
for the top Yukawa, can be read off Eq.\eqref{eq:qcoup} and
Eq.\eqref{eq:lcoup}. Following the field definition in
Eq.\eqref{eq:flavonfield} we use the notation
\begin{align}
g_{ij} \equiv 
g_{sf_{iL}f_{jR}} = i \, g_{af_{iL}f_{jR}} \; ,
\label{eq:scalarcoup}
\end{align}
to leading order and for all fermions except for $i = j = t$.

\subsubsection*{Flavon and top decays}

\begin{figure}[t]
\includegraphics[width=0.42\textwidth]{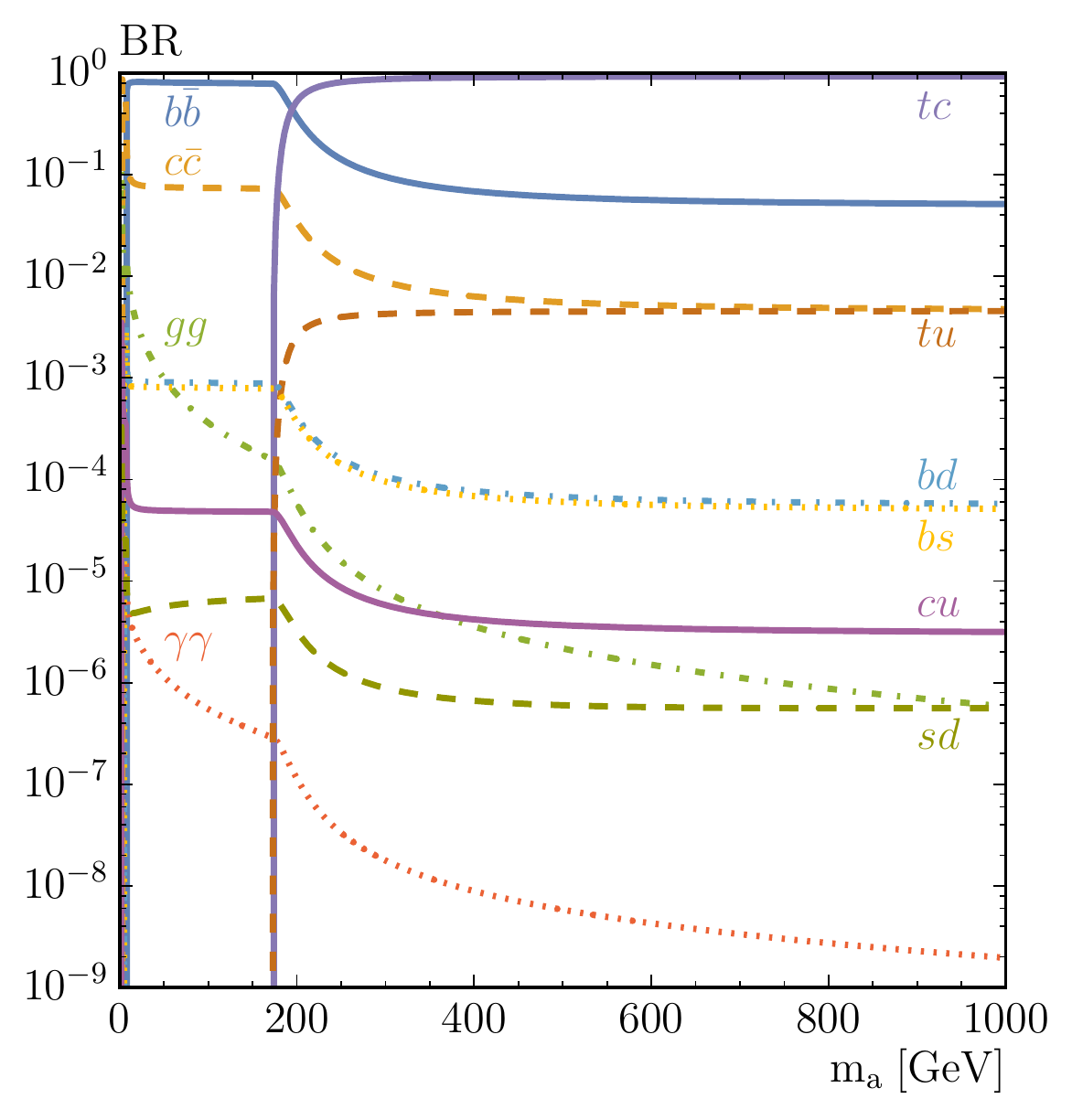}
\hspace*{0.10\textwidth}
\includegraphics[width=0.42\textwidth]{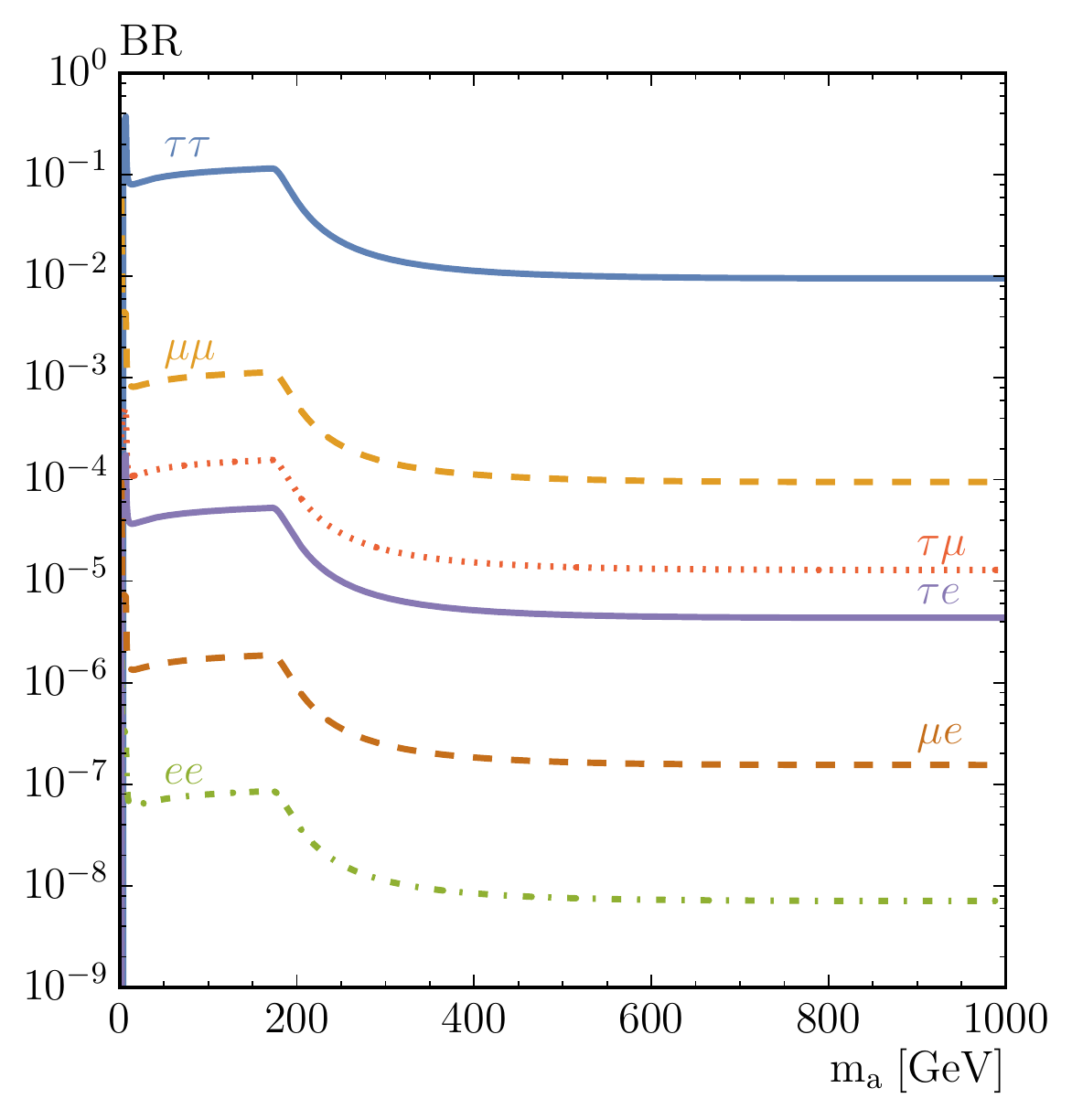}
  \caption{Flavon branching ratios for decays to quarks (left) and
    leptons (right).}
\label{fig:br}
\end{figure}

In terms of these flavon couplings to fermions we can compute the flavon
branching ratios, which will guide us to possible signatures at
colliders. Obviously, flavon decays to a pair of fermions occur at
tree level, but unlike for example a Higgs boson the decays do not
have to be flavor-diagonal. The general form of the corresponding
partial width is
\begin{align}
    \frac{\Gamma(a \to f_i \bar{f}_j)}{m_a} = \, & 
    \frac{N_c}{16 \pi} 
    \left[ \frac{(m_a^2 - (m_i + m_j)^2)
           (m_a^2 - (m_i - m_j)^2)}{m_a^4}
    \right]^{1/2}\\ \notag
   &\left[ 
           \left( |g_{ij}|^2+|g_{ji}|^2 \right)
           \left(1 - \frac{m_i^2+m_j^2}{m_a^2} \right) 
       - 2 \left( g_{ij} g_{ji} + g^*_{ij} g^*_{ji}\right)
           \frac{m_i m_j}{m_a^2}
    \right] \; .
\end{align}
In addition, we can compute the loop-induced partial widths to gluons
or photons in complete analogy to the Higgs. The numerical results for our
parameter choice $\eps = 0.23$ are given in Fig.~\ref{fig:br}. As long
as $m_a < m_t$ the main decay channels are similar to the Higgs, with
$a \to b\bar{b}$ dominating over $a \to \tau \tau$ due to the larger
Yukawa coupling and the color factor $N_c$. Above the top threshold
almost all pseudoscalar flavons decay to
\begin{align}
a \to tj + \bar{t}j 
\qqquad \text{with}\qquad 
\frac{\Gamma(a \to t\bar{u})}
     {\Gamma(a \to t\bar{c})} \approx \eps^2 \approx \frac{1}{20} \; .
\label{eq:decay}
\end{align}
The one obvious question for colliders searches will be if the charm
in the final state could be tagged to improve a top+jet resonance
signal. In our analysis we do not employ charm tagging and instead
leave it as an obvious experimentally driven improvement. Following the
construction of the Lagrangian without a suppression $f/\Lambda$ in
the top Yukawa, the diagonal decay $a \to t \bar{t}$ does not occur at
tree level.\bigskip

We can turn around the above discussion, which lead to the dominant
flavor-violating flavon decay shown in Eq.\eqref{eq:decay}: according
to Eq.\eqref{eq:qcoup} the couplings $g_{tc} \sim g_{ct}$ scale like $\eps^2
m_t/f \approx m_t/(20f)$. In the limit $m_c \ll m_a <
m_t$, the corresponding flavor-changing top decay width is given by
\begin{align}
\frac{\Gamma(t\to c a)}{m_t} = 
\frac{1}{32 \pi}
\left( |g_{ct}|^2+|g_{tc}|^2\right) \;
\left(1-\frac{m_a^2}{m_t^2}\right)^2 \;.
\label{eq:topdecay}
\end{align}
Anomalous top decays into Higgs and charm final states have been
searched for at LHC and will be discussed in detail in
Sec.~\ref{sec:coll}.

\section{Quark flavor constraints}
\label{sec:flavor}

To date, the most constraining measurements on our flavon model arise
from quark flavor physics. As usual, loop-induced meson mixing and
rare decays have the largest impact on our model, parameterized by the
flavon mass $m_a$, the VEV $f$, and the quartic coupling $\lambda_S$.

\subsubsection*{Neutral meson mixing}

\begin{figure}[b!]
  \includegraphics[width=.42\textwidth]{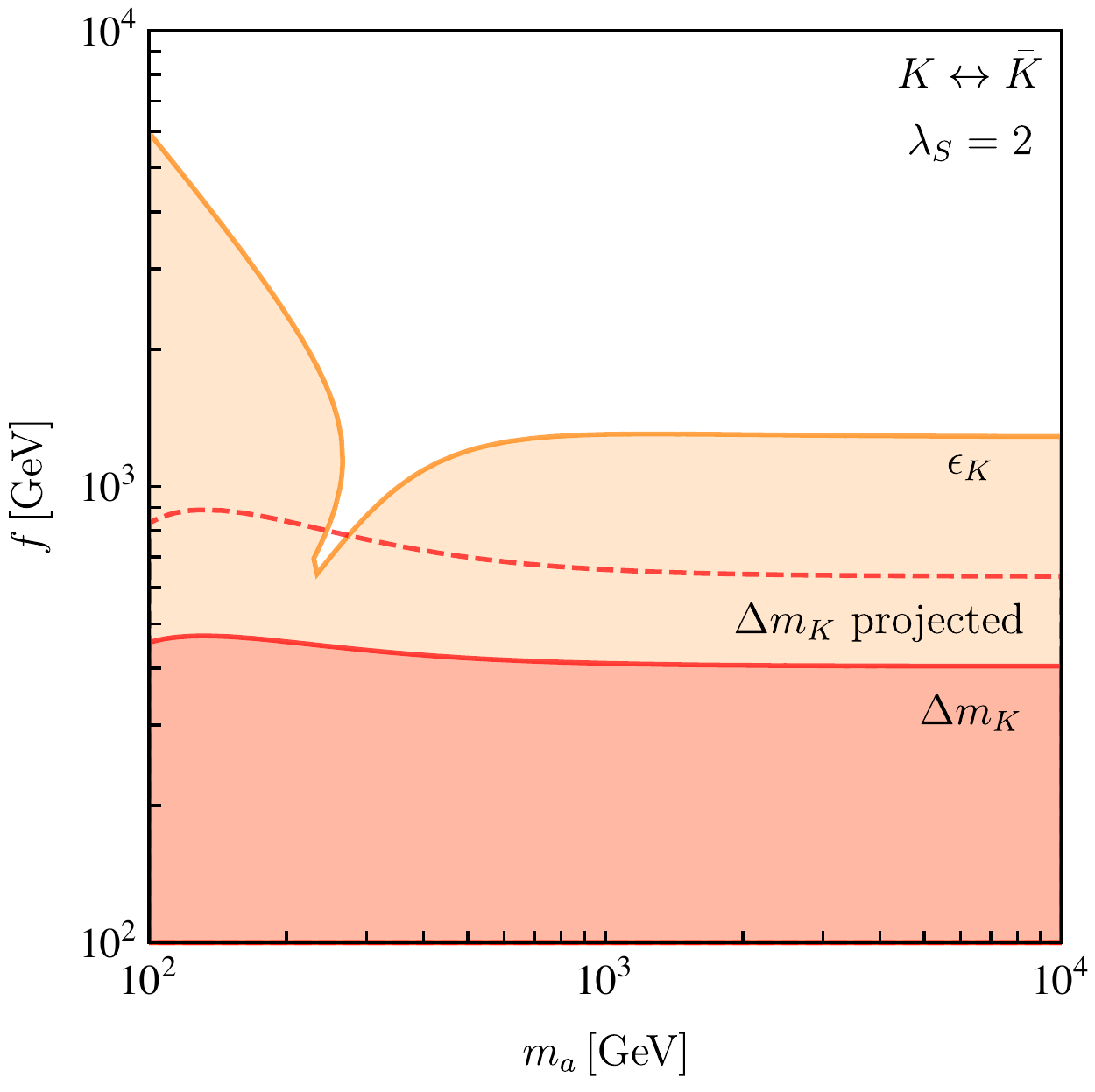}
  \hspace*{0.10\textwidth}
  \includegraphics[width=.42\textwidth]{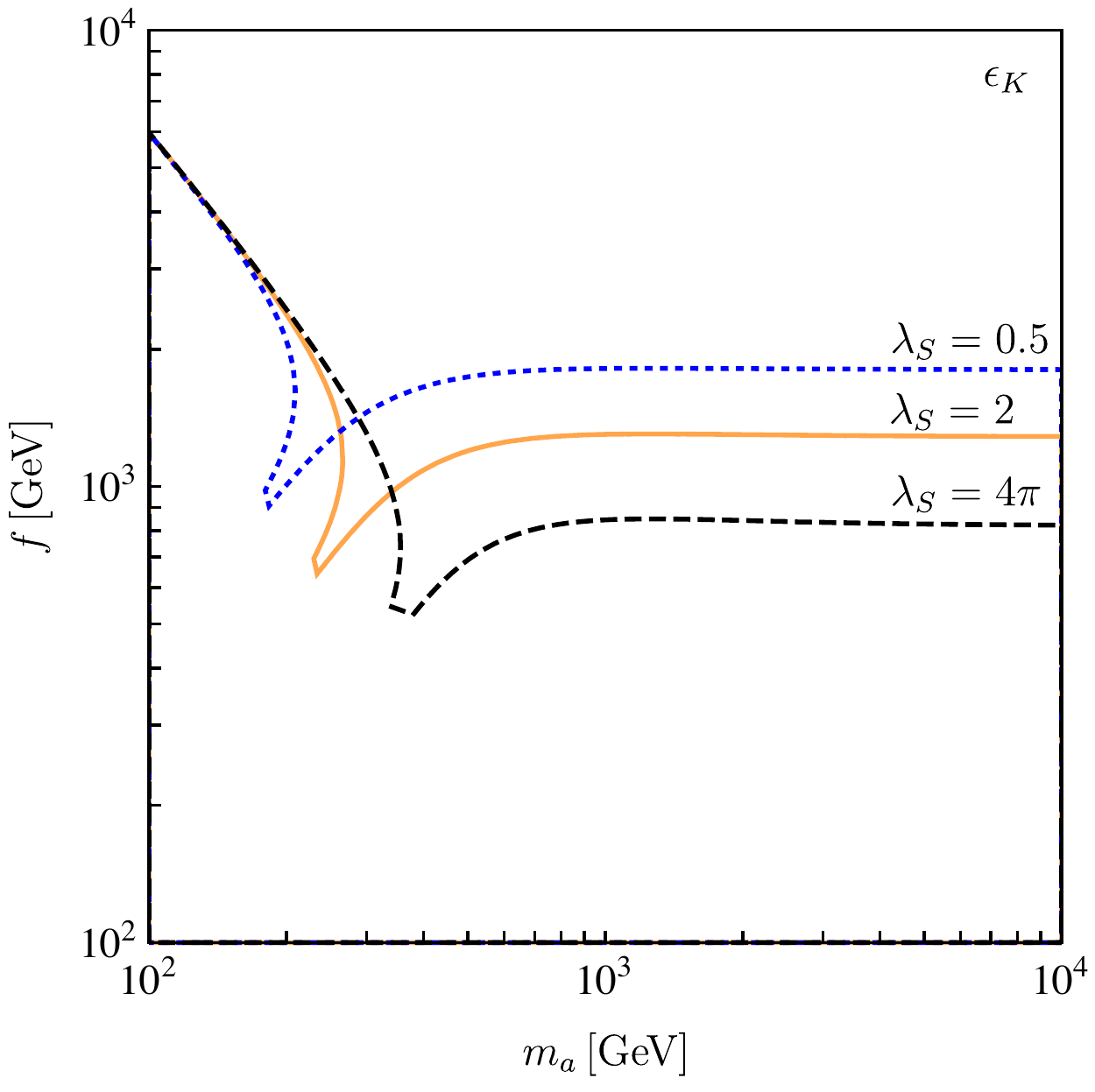}
  \caption{Left: regions in the $m_a-f$ plane excluded by flavon
    contributions to $\epsilon_K$ (orange) and $\Delta m_K$ (red) for
    our benchmark point and $\lambda_S=2$. The dashed red contour
    corresponds to the excluded region based on projected 
    improvements in $\Delta m_K$. Right: constraint from $\epsilon_K$
    for $\lambda_S=0.5$ (dotted blue), $\lambda_S=2$ (orange) and
    $\lambda_S =4\pi$ (dashed black). }
\label{fig:KK} 
\end{figure}

Because flavon models with the coupling structure given in
Eq.\eqref{eq:qcoup} lead to flavor-changing neutral currents, strong
limits are expected from meson anti-meson mixing. The effective
Hamiltonian describing $\Delta F =2$ interactions reads
\begin{align}
\Heff^{\Delta F=2}&=C_1^{ij} \,( \bar q^i_L\,\gamma_\mu \, q^j_L)^2+\widetilde C_1^{ij} \,( \bar q^i_R\,\gamma_\mu \, 
q^j_R)^2 +C_2^{ij} \,( \bar q^i_R \, q^j_L)^2+\widetilde C_2^{ij} \,( \bar q^i_L \, q^j_R)^2\notag\\
&+ C_4^{ij}\, ( \bar q^i_R \, q^j_L)\, ( \bar q^i_L \, q^j_R)\,+C_5^{ij}\, ( \bar q^i_L \,\gamma_\mu\, q^j_L)\, ( \bar q^i_R \,
\gamma^\mu q^j_R)\,+ \text{h.c.} 
\label{eq:heffdf2}
\end{align}
At tree-level, flavon exchange generates the Wilson
coefficients~\cite{Buras:2013rqa,Crivellin:2013wna}
\begin{align}
C_2^{ij} &= -(g_{ji}^*)^2\left(\frac{1}{m_s^2}-\frac{1}{m_a^2}\right)\notag \\
\tilde C_2^{ij} &= -g_{ij}^2\left(\frac{1}{m_s^2}-\frac{1}{m_a^2}\right)\notag \\
C_4^{ij} &= -\frac{g_{ij}g_{ji}}{2}\left(\frac{1}{m_s^2}+\frac{1}{m_a^2}\right)\,.
\label{eq:wilsons}
\end{align}
For $m_a=m_s$ the two contributions to $C_2$ and $\widetilde C_2$
cancel, while there is a constructive interference in $C_4$. Given
that the masses in Eq.\eqref{eq:masses} are set by independent scales,
such a cancellation would be accidental.  Depending on the meson
system, there can be sizable enhancement from RG running and matrix
elements. We implement RG running according to
Refs.~\cite{Bona:2007vi, Gorbahn:2009pp} with the matrix elements
given in Refs.~\cite{Carrasco:2015pra}, matching the
scalar and pseudoscalar flavon contributions at $\mu=m_s$ and
$\mu=m_a$, respectively. Fits based on projections of future
experimental improvements on meson mixing observables from LHCb and
Belle~II, as well as projected lattice improvements are collected in
Ref.~\cite{Charles:2013aka}.\bigskip

We start with 95\%~CL limits from $K-\bar K$
mixing~\cite{Bona:2007vi}
\begin{align}
C_{\eps_K}&=\frac{ \text{Im} \langle K^0|\mathcal{H}^{\Delta F=2}|\bar K^0\rangle}{\text{Im} \langle K^0| \mathcal{H}_\text{SM}^{\Delta F=2} |\bar K^0 \rangle} = 1.05_{-0.28}^{+0.36} \notag \\
C_{\Delta m_K}&=\frac{\text{Re}\langle K^0|\mathcal{H}^{\Delta F=2}|\bar K^0\rangle}{\text{Re} \langle K^0| \mathcal{H}_\text{SM}^{\Delta F=2} |\bar K^0 \rangle} = 0.93_{-0.42}^{+1.14} \; .
\end{align}
where $\mathcal{H}^{\Delta F=2}$ includes the SM and flavon
contributions, while $\mathcal{H}_\text{SM}^{\Delta F=2}$
parameterizes the SM contribution.  In the left panel of
Fig.~\ref{fig:KK} we show the region excluded by contributions from
scalar and pseudoscalar flavon exchange to $C_{\epsilon_K}$ and
$C_{\Delta m_K}$. The dip feature is due to the accidental
cancellation in $C_2^{sd}$ and $\tilde C_2^{sd}$, as shown in
Eq.\eqref{eq:wilsons}. It is a universal feature in $K-\bar K$ mixing,
unless the contribution to $C_4^{sd}$ completely dominates. For our
benchmark point, the dip in $C_{\Delta m_K}$ is below $m_a=100$~GeV,
not visible in the plot. The position also depends on the scalar
quartic $\lambda_S$, which also determines the excluded value of $f$
for large $m_a$. The dashed red contour corresponds to the excluded
region based on projected improvements in $\Delta m_K$, under
the optimistic assumptions presented in
Ref.~\cite{Charles:2013aka}. The right panel of Figure \ref{fig:KK}
shows the variation in the $C_{\epsilon_K}$ exclusion contour for
$\lambda_S=0.5, 2, 4\pi$.\bigskip

\begin{figure}[t]
  \includegraphics[width=.42\textwidth]{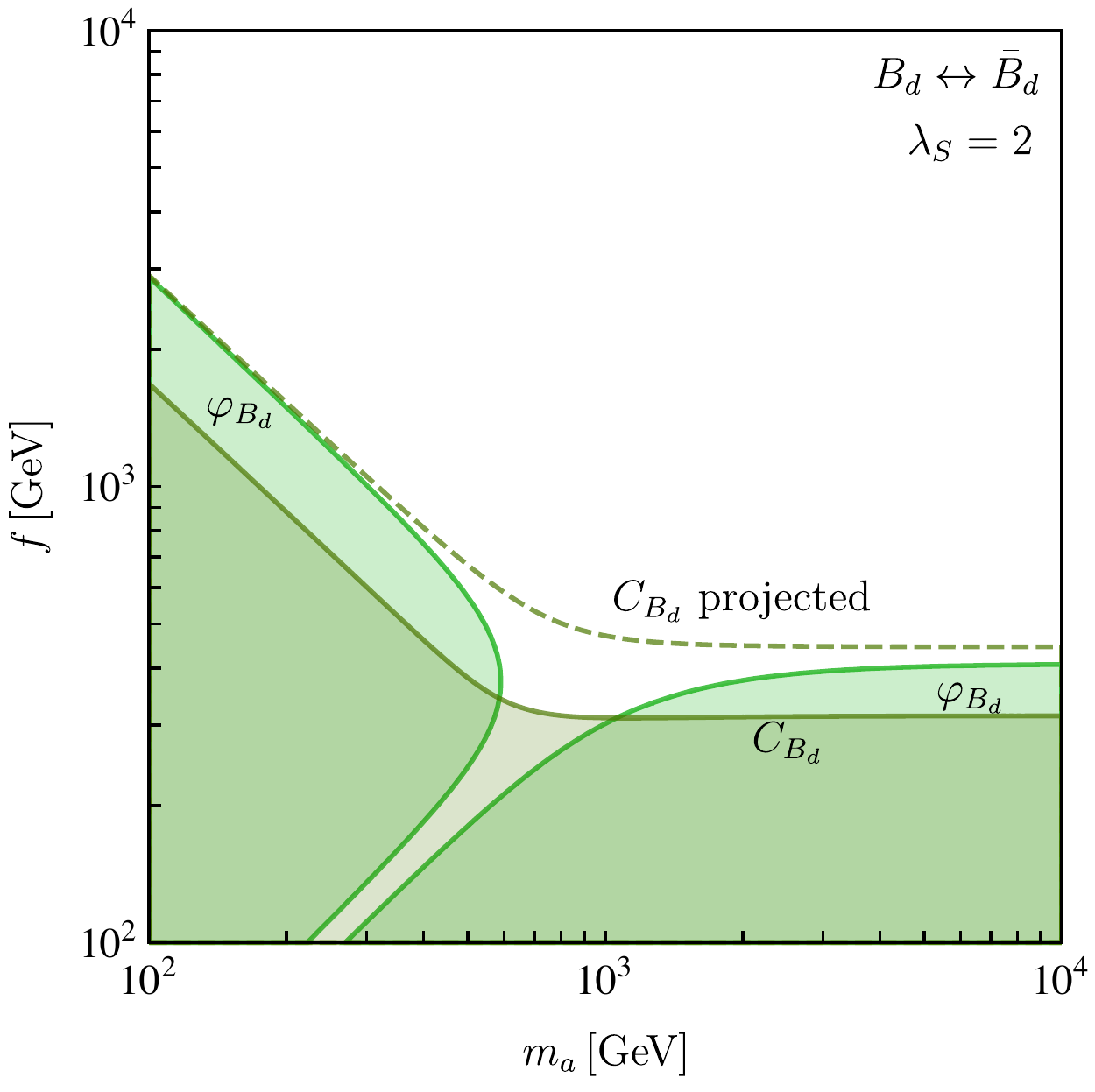}
  \hspace*{0.10\textwidth}
  \includegraphics[width=.42\textwidth]{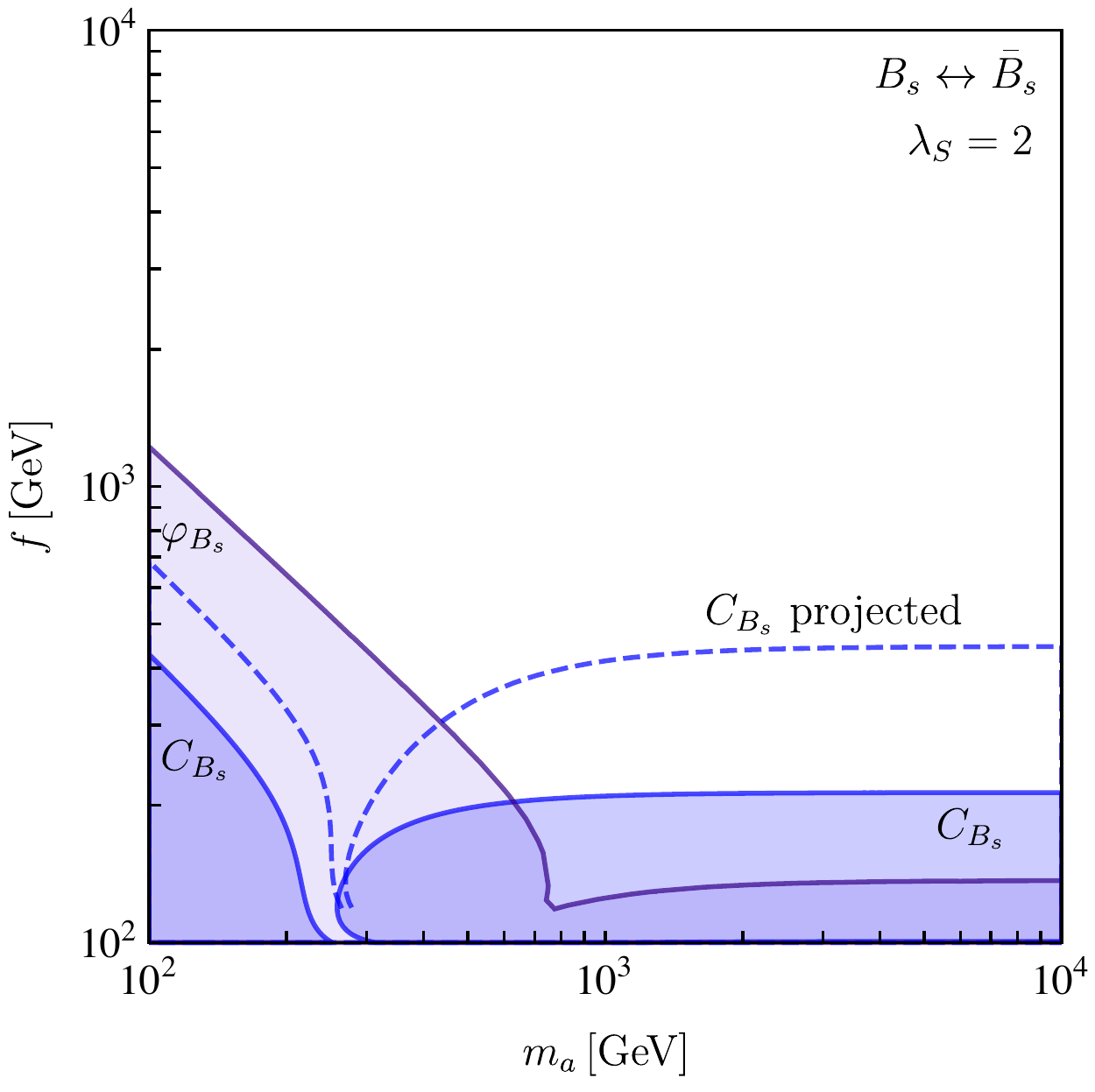}
  \caption{Left: regions in the $m_a-f$ plane excluded by flavon
    contributions to $C_{B_d}$ (light green) and $\varphi_{B_d}$
    (green) for our benchmark point and $\lambda_S=2$. Right:
    constraints from flavon contributions to $C_{B_s}$ (blue) and
    $\varphi_{B_s}$ (light purple). The dashed contours correspond to
    the excluded regions based on projected improvements in
    $C_{B_d}$ and $C_{B_s}$.}
\label{fig:Bq} 
\end{figure}

For the two versions of $B -\bar B$ mixing we define
\begin{align}
C_{B_q}e^{2i\varphi_{B_q}}=\frac{\langle B_q|\mathcal{H}^{\Delta F=2}|\bar B_q\rangle}{\langle B_q| \mathcal{H}_\text{SM}^{\Delta F=2} |\bar B_q \rangle}\; ,
\end{align}
with the 95\%~CL limits~\cite{Bona:2007vi}
\begin{align}
C_{B_d}&= 1.07_{-0.31}^{+0.36}  & \varphi_{B_d}&= -2.0_{-6.0}^{+6.4} \notag \\
C_{B_s}&= 1.052_{-0.152}^{+0.178} & \varphi_{B_s}&= 0.72_{-2.28}^{+3.98} \; .
\end{align}
Figure~\ref{fig:Bq} shows the excluded regions in the $f-m_a$ plane
for our benchmark point.  The optimistic projected improvements
in $C_{B_d}$ and $C_{B_s}$ follow Ref.~\cite{Charles:2013aka}.\bigskip

\begin{figure}[t]
\begin{center}
  \includegraphics[width=.42\textwidth]{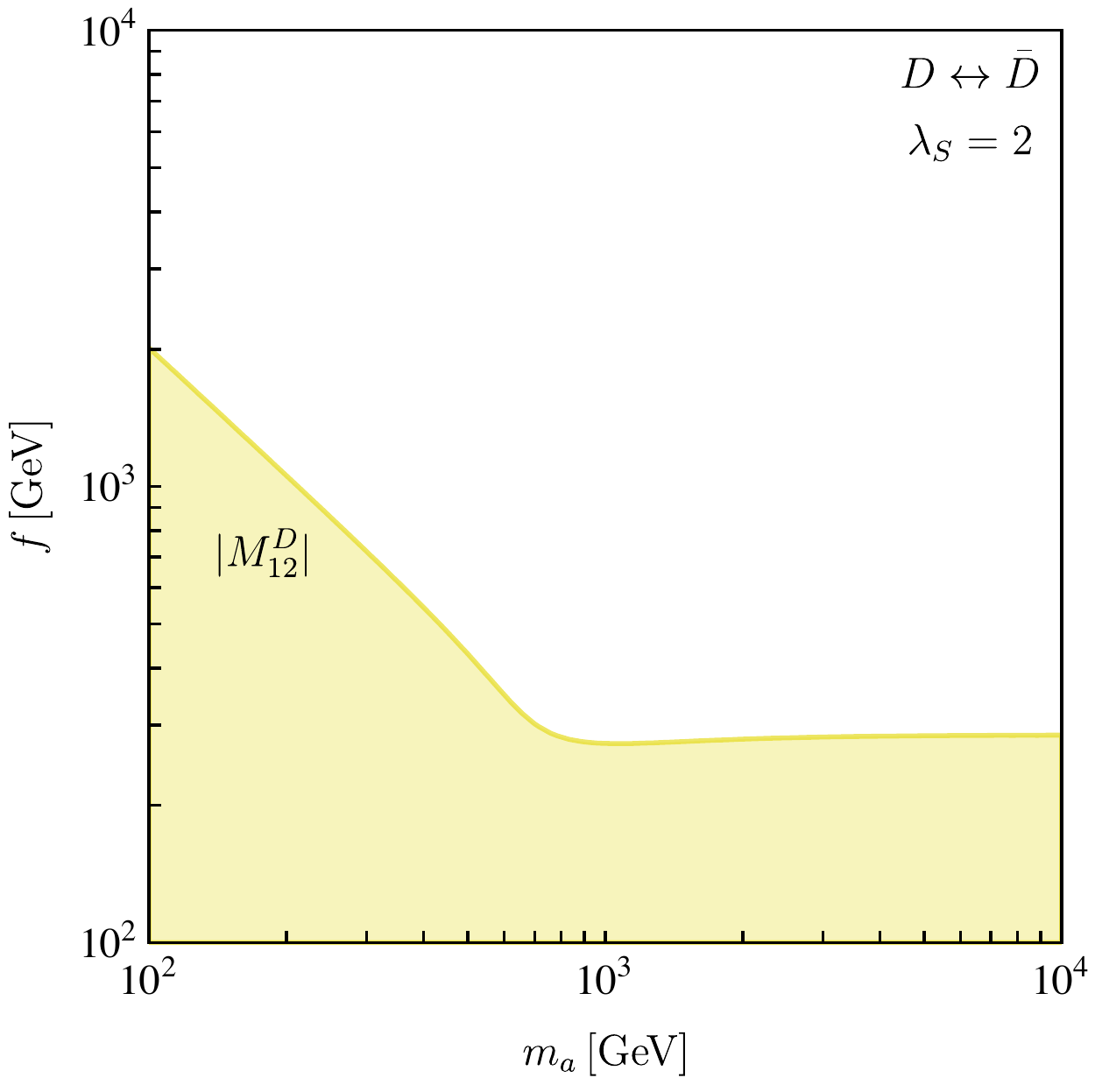}
\end{center}
  \caption{Regions in the $m_a-f$ plane excluded by flavon
    contributions to $|M_{12}^D|$ (shaded yellow) for our benchmark
    point and $\lambda_S=2$.}
  \label{fig:DD}
\end{figure}

Finally, since the SM contribution to $D-\bar D$ mixing is plagued
with very large hadronic uncertainties, we define and only demand that
the flavon contributions do not exceed the $2\sigma$
constraint~\cite{Bevan:2014tha}
\begin{align}
|M^{D}_{12}|
=|\langle D|\mathcal{H}^{\Delta F=2}|\bar D\rangle
< 7.7~\text{ps}^{-1} \; .
\end{align}
The results for our benchmark point are shown in Fig.~\ref{fig:DD}. In
principle, the sizable flavon coupling $g_{tc,ct}$ could result in
sizable loop contributions from one-loop box diagram.  Altogether, we
find a relative suppression of the kind $m_t^2\epsilon^2/(4\pi^2f^2)$ with
respect to the tree level diagram, which renders the loop
contributions completely negligible for the parameter space of
interest.

\subsubsection*{Leptonic meson decays}

\begin{figure}[t]
  \includegraphics[width=.42\textwidth]{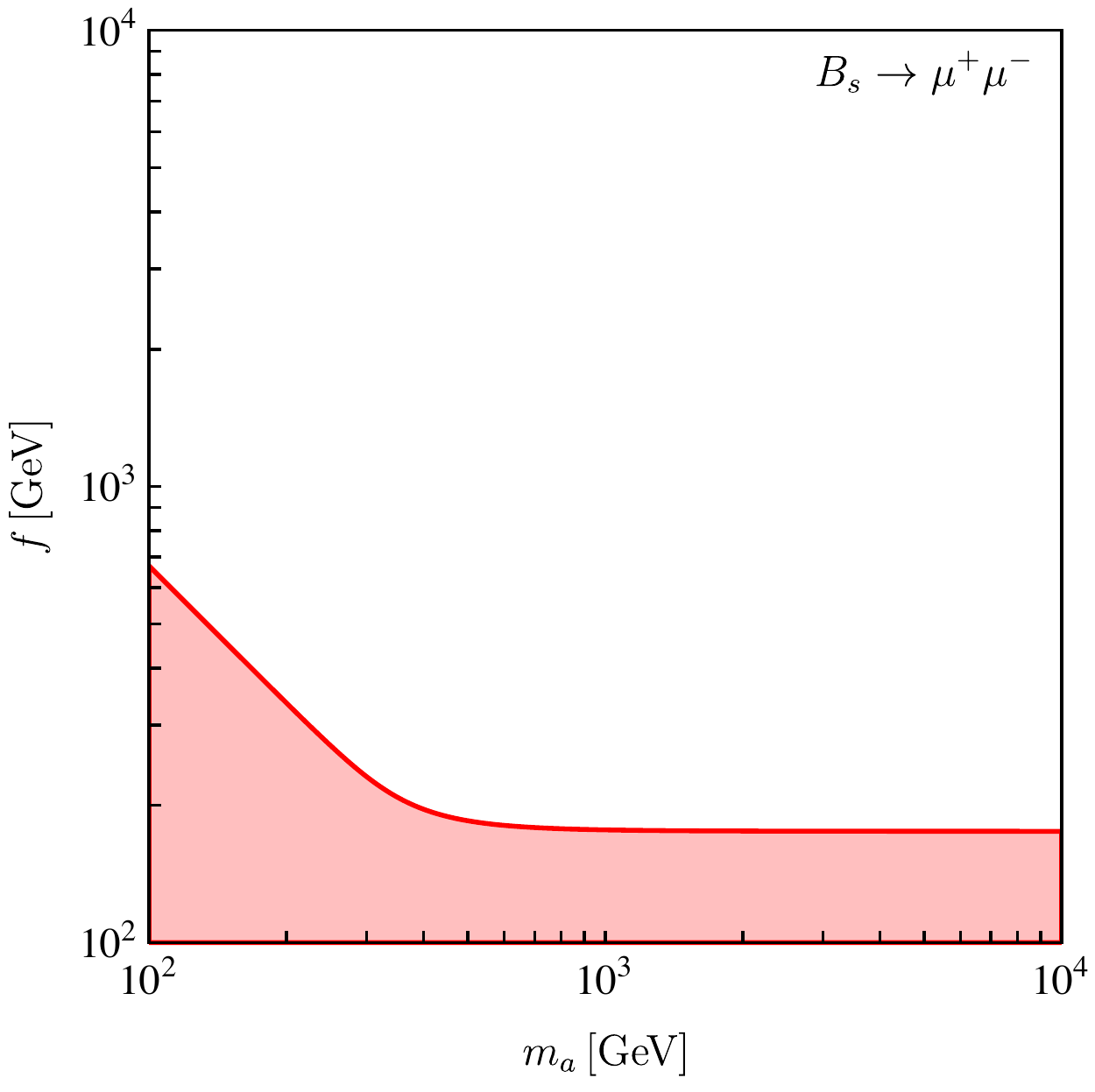}
  \hspace*{0.10\textwidth}
  \includegraphics[width=.42\textwidth]{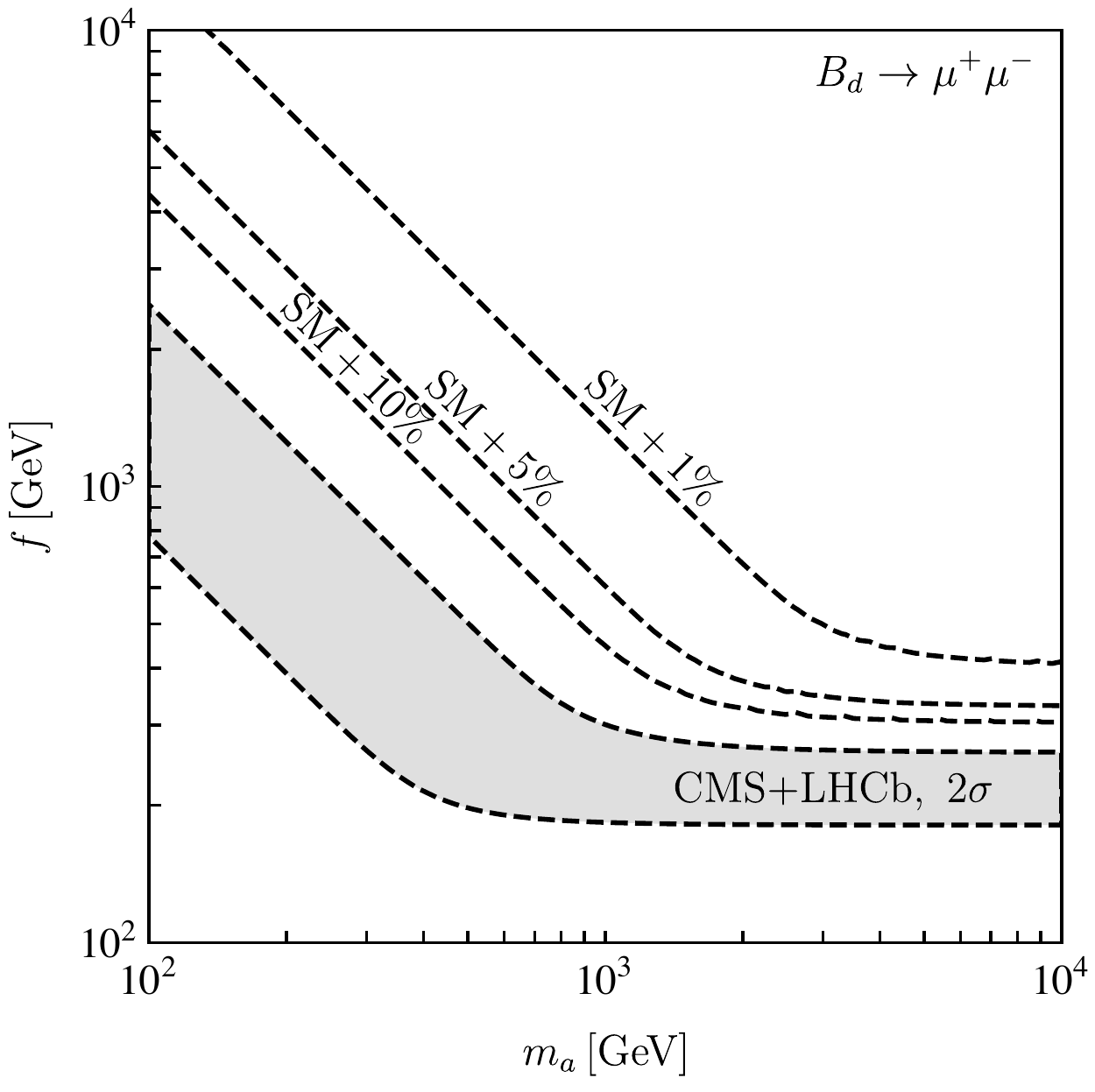}
  \caption{Left: regions in the $m_a-f$ plane excluded by flavon
    contributions to the decay $B_s\rightarrow \mu^+\mu^-$. Right:
    parameter space where the branching ratio for $B_d\rightarrow
    \mu^+\mu^-$ stays within the $2\sigma$ confidence interval (shaded
    gray), as well as contours of $1\%$, $5\%$ and $10\%$ enhancement
    with respect to the SM prediction.}
\label{fig:Bmumu} 
\end{figure}

Flavon-mediated decays of neutral mesons into charged leptons can be
described by the effective Hamiltonian
\begin{align}
\mathcal{H}_\text{eff}=-\frac{G_F^2m_W^2}{\pi^2}\,\left(C_S^{ij}\, (\bar q_iP_L q_j)\bar \ell \ell +\tilde  C_S^{ij}\, (\bar q_iP_R q_j)\bar \ell \ell + C_P^{ij}(\bar q_iP_L q_j)\bar \ell \gamma_5 \ell+\tilde C_P^{ij}(\bar q_iP_R q_j)\bar \ell \gamma_5 \ell \right)+ \text{h.c.} \; .
\end{align}
The branching ratio for the meson decay of a neutral meson is given by
\begin{align}
\br(M\rightarrow \ell^+ \ell^- )=&
\frac{G_F^4m_W^4}{8\pi^5} \beta \,m_M f_M^2 m_\ell^2 \tau_M \notag\\
&\left[ \left|\frac{m_M^2\big(C_P^{ij}-\tilde C^{ij}_P\big)}{2m_\ell (m_i+m_j)}-C_A^\text{SM}\right|^2+
\left|\frac{m_M^2\big(C_S^{ij}-\tilde C^{ij}_S\big)}{2m_\ell (m_i+m_j)}\right|^2\beta^2
\right] \; ,
\end{align}
where $\beta(x)=\sqrt{1-4x^2}$ with $x = m_\ell/m_M$.  As for meson
mixing, we only need to consider tree-level flavon contributions to
the corresponding Wilson coefficients,
\begin{align}
C_S^{ij}&=\frac{\pi^2}{2G_F^2 m_W^2}\frac{2g_{\ell\ell}g_{ji}}{m_s^2} 
&\tilde C_S^{ij}&=\frac{\pi^2}{2G_F^2 m_W^2}\frac{2g_{\ell\ell}g_{ij}}{m_s^2} \notag\\
C_P^{ij}&=\frac{\pi^2}{2G_F^2 m_W^2}\frac{2g_{\ell\ell}g_{ji}}{m_a^2} 
&\tilde C_P^{ij}&=\frac{\pi^2}{2G_F^2 m_W^2}\frac{2g_{\ell\ell}g_{ij}}{m_a^2}\; .
\end{align}
Since the scalar contributions do not interfere with the SM
contribution, the resulting constraints are almost independent from
the scalar mass. In addition, they are insensitive to the value of the
quartic coupling $\lambda_S$. In contrast, the SM contribution is
generated at one loop, and for the $B_s$-system is to a very good
approximation given by
\begin{align}
C_\text{SM}= -V_{tb}^*V_{ts}\,Y\left(\frac{m_t^2}{m_W^2}\right) \; ,
\label{eq:YY}
\end{align}
with 
\begin{align}
Y(x)=\eta_\text{QCD} \; \frac{x}{8}\left[\frac{4-x}{1-x}+\frac{3x}{(1-x)^2} \log x \right] \; ,
\end{align}
where $\eta_\text{QCD}=1.0113$ parametrizes higher order
corrections~\cite{Buras:2012ru}. Due to the sizable width difference
of the $B_s$-meson system, the theoretical prediction has to be
rescaled by $(1-y_s)^{-1}$, where
$y_s=0.088\pm0.014$~\cite{Fleischer:2012bu}, before being compared
with the experimental result~\cite{CMS:2014xfa}.
\begin{align}
\br(B_s\rightarrow \mu^+\mu^-)=2.8^{+0.7}_{-0.6}\e{-9}\,.
\end{align}
The corresponding limits on our flavon benchmark point are shown in
the left panel of Fig.~\ref{fig:Bmumu}.  In the case of the
$B_d$-system, the corresponding correction is negligible and the SM
prediction follows from a straightforward replacement of indices in
Eq.\eqref{eq:YY}. The recent combination of
CMS~\cite{Chatrchyan:2013bka} and LHCb~\cite{Aaij:2013aka}
measurements yields~\cite{CMS:2014xfa}
\begin{align}
\br(B_d\rightarrow \mu^+\mu^-)=(3.6\pm1.6) \e{-10}\,.
\end{align}
For this channel we require our flavon contributions to stay within
the $2\sigma$ interval, namely $\br(B_d\rightarrow \mu^+\mu^-)=
[1.4,7.4]\e{-10}$.  In the right panel of
Fig.~\ref{fig:Bmumu} we show where the flavon contributions agree
with measured value of $\br (B_d\rightarrow \mu^+\mu^-)$, as well as
$1\%$, $5\%$ and $10\%$ enhancements with respect to the SM
prediction. While an explanation of the $2\sigma$ deviation is in
tension with constraints from neutral meson mixing, flavon exchange
can lead to sizable enhancements.\bigskip

In addition to these bottom-mesons we can also derive constraints from
$D \rightarrow \mu^+\mu^-$ decays, which turn out considerably
weaker. Finally, flavon limits from $K_L \rightarrow \mu^+\mu^-$
decays exclude a region in parameter space very similar to the one
ruled out by $B_s\rightarrow \mu^+\mu^-$.

\section{Future lepton flavor measurements}
\label{sec:lepton}

While currently the experimental results from quark flavor physics are
most constraining for our flavon models, we discuss a set of upcoming
experiments in lepton flavor physics which will dramatically improve
in the coming years. We can use the same benchmark point for these
lepton flavor experiments as for collider searches, because the lepton
and quark sectors of our flavon models can be adapted independently.

\begin{figure}[b!]
  \includegraphics[width=.55\textwidth]{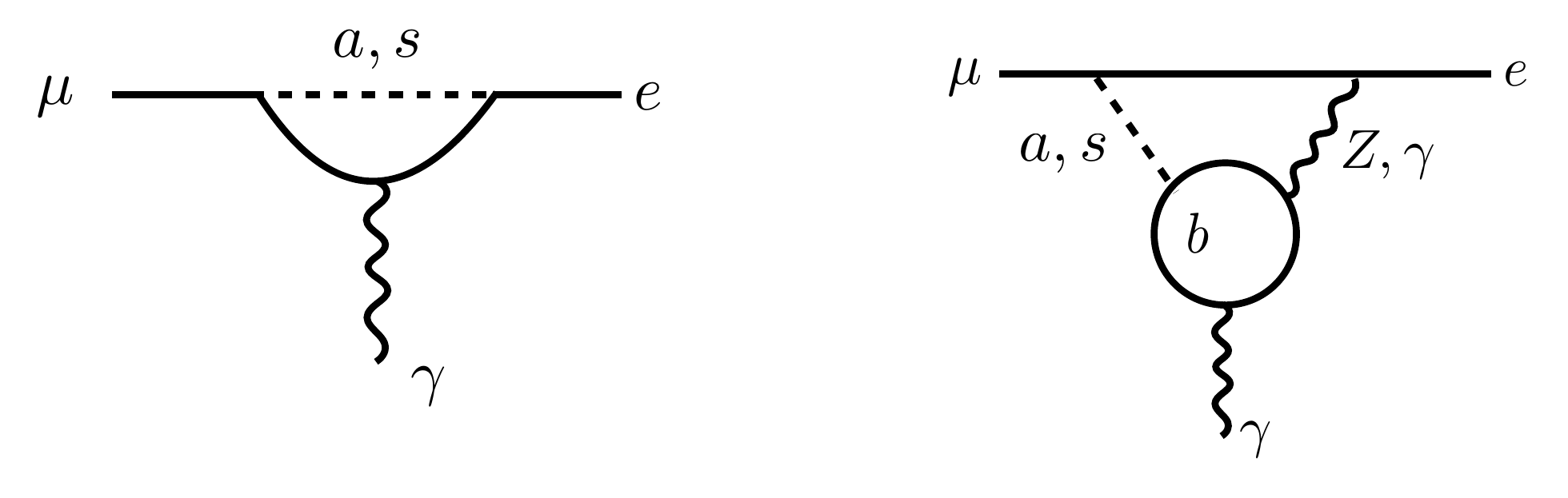}
  \hspace*{0.04\textwidth}
  \raisebox{0.6cm}{\includegraphics[width=.28\textwidth]{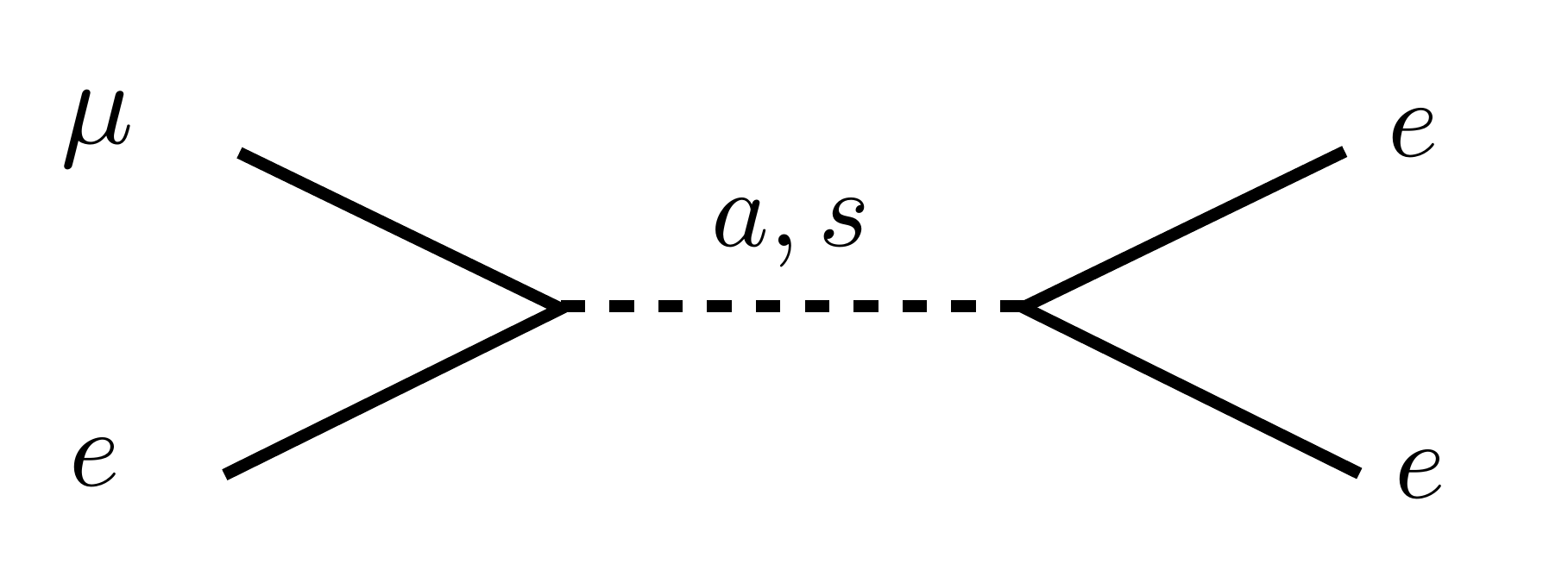}}
\caption{Feynman diagrams showing flavon contributions to $\mu
  \to e \gamma$ at one-loop level and two-loop level, as well as
  flavon contributions to $\mu \to 3e$.}
\label{fig:MuEG} 
\end{figure}

\subsubsection*{Decay $\mu\rightarrow e\gamma$ }

\begin{figure}[t]
  \centering
  \includegraphics[width=.42\textwidth]{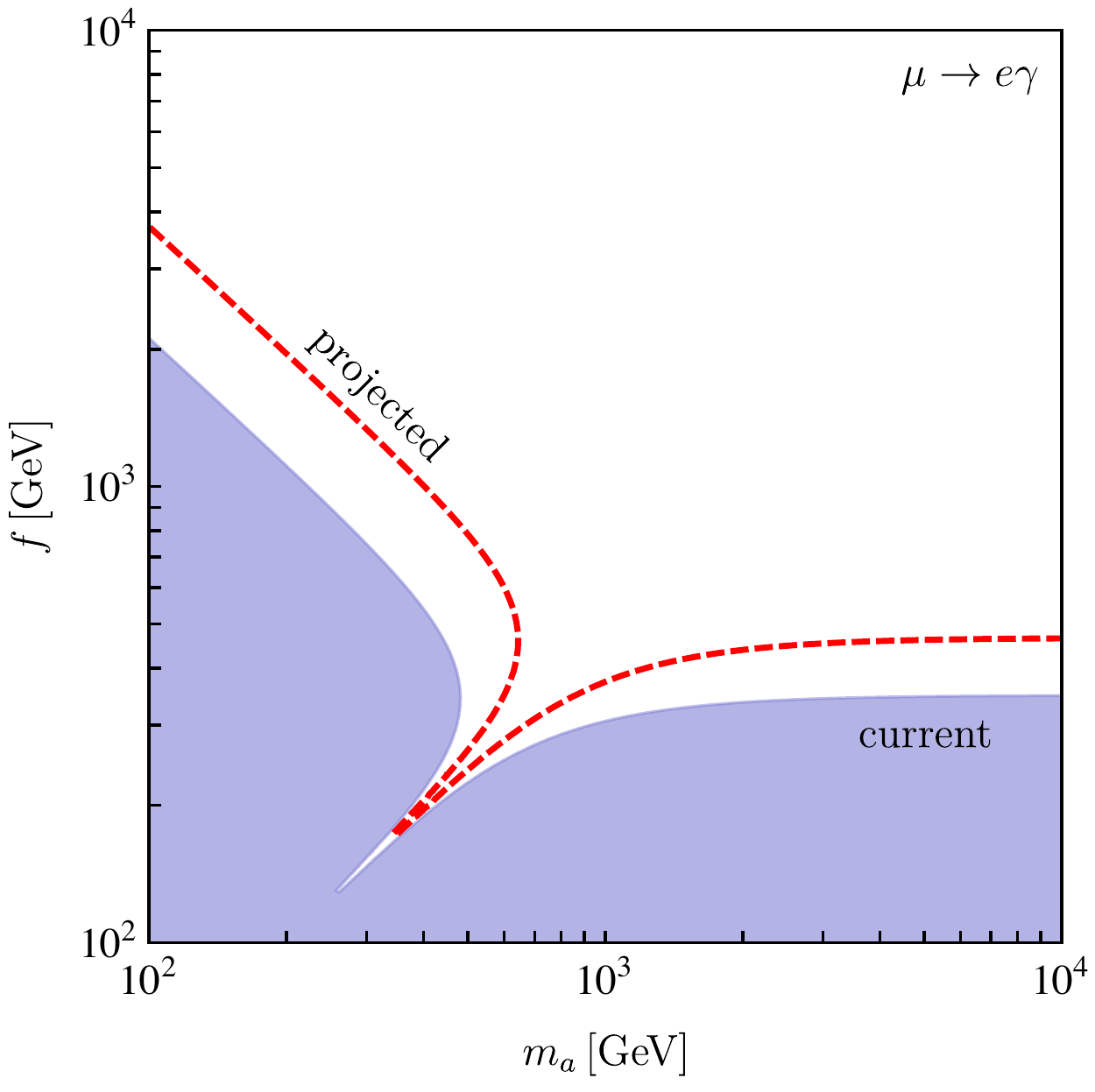}
  \caption{Regions in the $m_a-f$ plane excluded by flavon
    contributions to $\mu \rightarrow e \gamma$. The corresponding
    bounds from $\tau \rightarrow \mu \gamma$ and $\tau \rightarrow e
    \gamma$ are not visible for the plotted parameter range.}
\label{fig:MuEGbounds} 
\end{figure}

Radiative leptonic decays are mediated by dipole operators
\begin{align}
\lag_{\text{eff}}&=m_{\ell'}\, C_T^L\,\bar \ell \sigma^{\rho\lambda}P_L\,\ell' \,F_{\rho\lambda}+m_{\ell'}\, C_T^R\,\bar \ell \sigma^{\rho\lambda}P_R\,\ell'\,F_{\rho\lambda}\,.
\label{eq:MueGamma}
\end{align}
giving a branching ratio
\begin{align}
\br(\ell'\rightarrow  \ell\gamma)=\frac{m_{\ell'}^5}{4\pi \Gamma_{\ell'}}\left(|C_T^L|^2+|C_T^R|^2\right) \; .
\end{align}
The relevant one-loop diagram for the flavon contribution, shown in
Fig.~\ref{fig:MuEG}, gives the Wilson coefficients
\begin{align}
C_T^L = (C_T^R)^*
=\frac{g}{32\pi^2}\sum_{k=e,\mu,\tau} &\bigg\{ \frac{1}{6}\left( \,g^*_{\ell k}g_{\ell' k}+\frac{m_\ell}{m_k}g^*_{k \ell }g_{k\ell' }\right)\left(\frac{1}{m_s^2}-
\frac{1}{m_a^2}\right)\notag\\
&-g_{\ell k}g_{k\ell'}\frac{m_k}{m_{\ell'}}
\left[ \frac{1}{m_s^2}
\left(\frac{3}{2}+\log\frac{m_{\ell'}^2}{m_s^2}\right)-\frac{1}{m_a^2}
\left(\frac{3}{2}+
\log\frac{m_{\ell'}^2}{m_a^2}
\right)\right]
\bigg\} \; .
\label{eq:muegamW}
\end{align}
In particular for $\mu \to e \gamma$ the chirally enhanced second term
in Eq.\eqref{eq:muegamW} leads to sizable contributions. Current
experimental bounds are~\cite{Adam:2013mnn,Hayasaka:2007vc} 
\begin{align}
\br(\mu \rightarrow e \gamma) < 5.7\e{-13} 
\qquad \text{and} \qquad 
\br(\tau \rightarrow \mu \gamma) < 4.5\e{-8}\; ,
\end{align}
while the upgraded MEG~II experiment has a projected sensitivity
of~\cite{Baldini:2013ke}
\begin{align}
\br(\mu \rightarrow e \gamma)=6\e{-14} \; .
\end{align}
In Fig.~\ref{fig:MuEGbounds} we show the different constraints on our
flavon model. 

\subsubsection*{Conversion $\mu\rightarrow e$}

\begin{figure}[b!]
\centering
  \includegraphics[width=.45\textwidth]{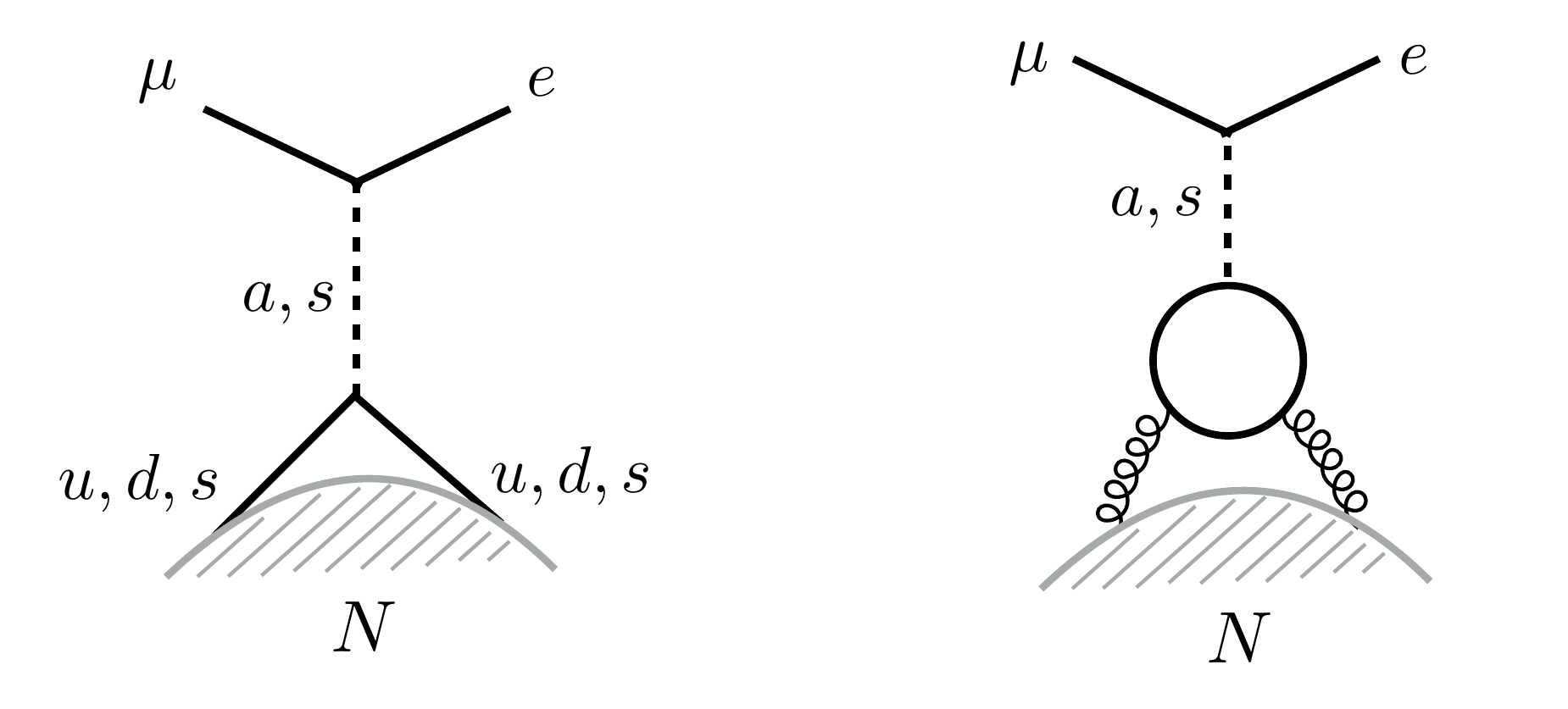}
\caption{Diagrams showing flavon contributions to
  $\mu \to e $ conversion in nuclei at tree level and one-loop
  level.}
\label{fig:MuToE} 
\end{figure}

\begin{figure}[t]
  \centering
  \includegraphics[width=.42\textwidth]{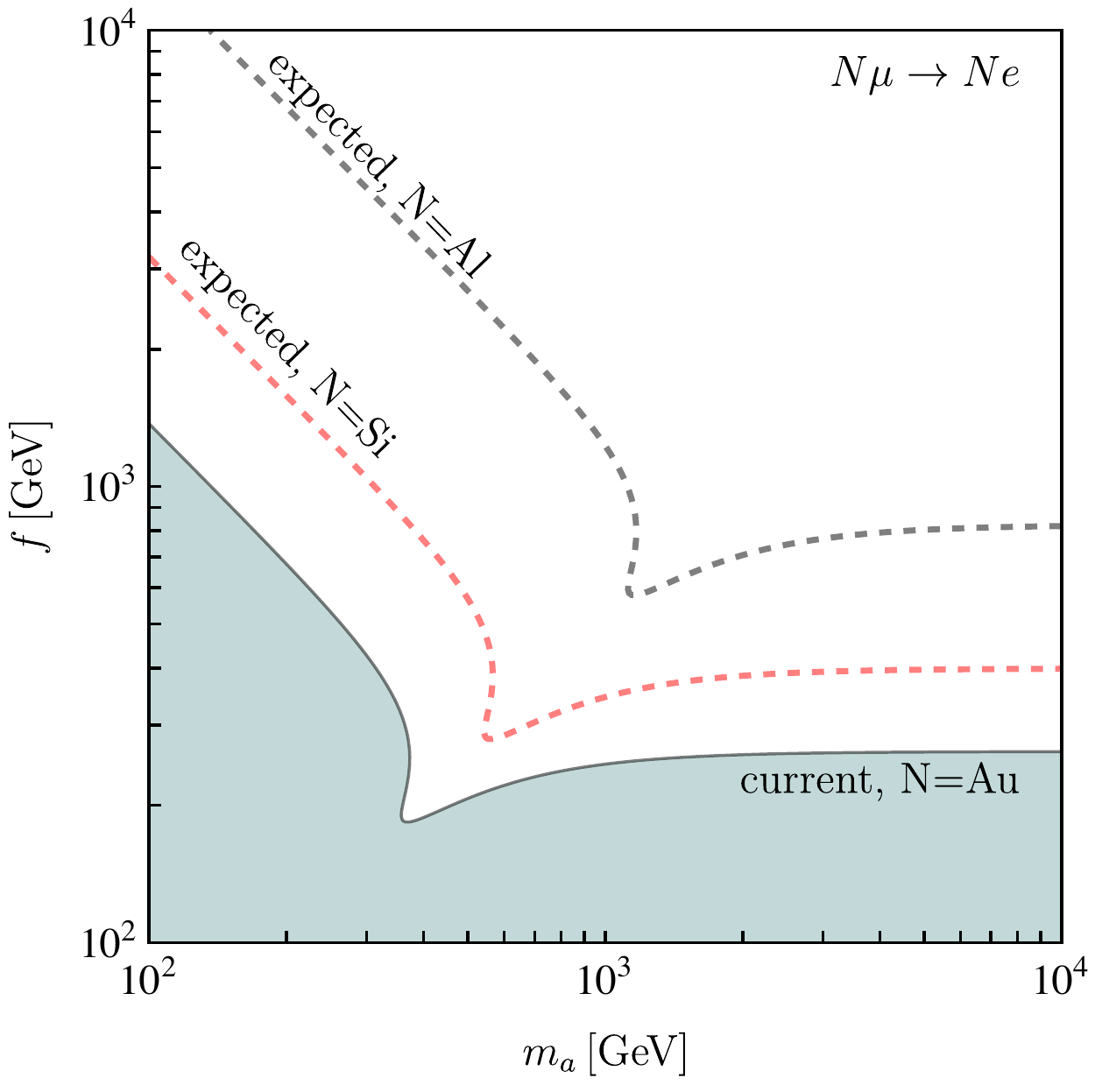}
  \caption{Regions in the $m_a-f$ plane excluded by flavon
    contributions to the conversion $N \mu \rightarrow N e$.}
\label{fig:MuEConvbounds} 
\end{figure}

In addition to the dipole operators shown in Eq.\eqref{eq:MueGamma},
the following effective operators contribute to $N\mu\rightarrow N e$
conversion
\begin{align}
\lag_{\text{eff}}=
C_{qq}^{VL}\,\bar e \gamma^\nu P_L \mu\, \bar q \gamma_\nu q
+m_\mu m_q\,C_{qq}^{SL}\bar e P_R \mu \,  \bar q q
+m_\mu \alpha_s C_{gg}^L\,\bar e P_R \mu \,G_{\rho\nu}G^{\rho\nu}\,+ (R\leftrightarrow L) \; ,
\label{eq:ConvLag}
\end{align}
Before we include the nuclear effects to compute the actual conversion
rate, we derive the Wilson coefficients induced by flavon
exchange. The relevant diagram is shown on the left of
Fig.~\ref{fig:MuToE} and gives us
\begin{align}
C^{SL}_{qq}&=\left(\frac{1}{m_s^2}+\frac{1}{m_a^2}\right)g_{\mu e}^*\text{Re}(g_{qq})\,,\notag\\
C^{SR}_{qq}&=\left(\frac{1}{m_s^2}-\frac{1}{m_a^2}\right)g_{ e\mu} \text{Re}(g_{qq})\,.
\end{align}
%
Contributions to $C_{gg}^{L,R}$ arise only from integrating out the
non-dynamical heavy quarks and we absorb them in $\tilde C_p^{SL}$ and
$\tilde C_n^{SL}$. The relevant diagram is shown on the right hand
side of Fig.~\ref{fig:MuToE}. We further confirm that contributions
from vector operators are smaller than all scalar Wilson coefficients
and can be neglected~\cite{Harnik:2012pb}.  Barr-Zee-type diagrams, as
shown in Fig.~\ref{fig:MuEG}, which generate the dominant
contributions to both $\mu\to e \gamma$ and $\mu \to e$ conversion for
lepton flavor violating Higgs couplings are small due to the absence
of couplings to the top quark.

Next, we need to account for the effects of quarks inside the nucleons. We define
the nucleon-level Wilson coefficients 
\begin{align}
\tilde C_p^{VL} =\sum_{q=u,d} C^{VL}_{qq}\, f_{V_q}^p 
\qquad \text{and} \qquad 
\tilde C_p^{SL} =\sum_{q=u,d,s} C^{SL}_{qq}\, f_{q}^p-\sum_{Q=c,b,t} C^{SL}_{QQ} \,f_\text{heavy}^p\,,
\end{align}
in which $f_{V_q}^{p}, f_q^p$, and
$f_\text{heavy}^p=2/27\big(1-f_u^p-f_d^p-f_s^p\big)$ account for the
quark content of the proton~\cite{Shifman:1978zn}. Analogous
expressions hold for the neutron.  We use the numbers given in
Refs.\cite{Crivellin:2013ipa, Crivellin:2014cta}, based on the lattice
average from Ref.~\cite{Junnarkar:2013ac},
\begin{align}
f_u^p&=0.0191 \qqquad  f_u^n=0.0171\,,\notag\\
f_d^p&=0.0363 \qqquad f_d^n=0.0404\,,\notag\\
f_s^p&=f_s^n=0.043\,.
\end{align}
Using the $\sigma$-term derived from $SU(3)_C$ relations does not
change the results qualitatively.  Finally, we can compute the
conversion rate including effects from the nucleus' structure, 
\begin{align}
\Gamma_{N\mu\rightarrow N e}=\frac{m_\mu^5}{4}\left|C_T^L D +4\left[m_\mu m_p\tilde C_p^{SL}+\tilde C_p^{VL}V^p+ (p\rightarrow n) \right]\right|^2\,,
\label{eq:Convrate}
\end{align}
with $p$ and $n$ denoting the proton and neutron, respectively.  The
coefficients $D, S^{p,n}$ and $V^{p,n}$ are dimensionless functions of
the overlap integrals of the initial state muon and the final-state
electron wave-functions with the target nucleus. We
use the numerical values~\cite{Kitano:2002mt}
%
\begin{center}
\begin{tabular}{p{2.0cm}|P{1.5cm} P{1.5cm} P{1.5cm} P{1.5cm} P{1.5cm} P{2cm}}
\hline
\text{Target}& $D$& $S^p$ & $S^n$ & $V^p$& $V^n$&$\Gamma_\text{capt} [10^{-6} \text{s}]$\\\hline
\text{Au}&  0.189& 0.0614&0.0918&0.0974&0.146&13.06\\
\text{Al}&  0.0362& 0.0155&0.0167&0.0161&0.0173&0.705\\
\text{Si}&  0.0419& 0.0179&0.0179&0.0187&0.0187&0.871 \\ \hline
\end{tabular}
\end{center}
%
with $\Gamma_\text{capt}$ denoting the muon capture rate.\bigskip

Currently, the strongest experimental bound on $\mu \rightarrow e$ conversion is
set by SINDRUM~II, using a gold target~\cite{sindrum_now}
\begin{align}
\br(\mu\rightarrow e)^\text{Au}< 7\e{-13}\,;
\end{align}
but the future DeeMe~\cite{Natori:2014yba} and
COMET~\cite{Kuno:2013mha} experiments as well as
Mu2e~\cite{Abrams:2012er} aim to improve these bounds using a silicon
or an aluminum target. Their projections are
\begin{align}
\br(\mu\rightarrow e)^\text{Si}< 2\e{-14} 
\qquad \text{and} \qquad 
\br(\mu\rightarrow e)^\text{Al}< 6\e{-17} \; .
\end{align}
The region excluded by the current and future limits are shown in
Fig.~\ref{fig:MuEConvbounds}.  Compared to the quark flavor
constraints for example from meson mixing we see that current lepton
flavor constraints are weaker, but will soon become dominant.

\subsubsection*{Decays $\mu\rightarrow 3e$ and $\tau \rightarrow 3 \mu$}

Finally, we can exploit decays similar to $\mu \to e \gamma$, but
including weak boson effects.  The effective Lagrangian
parametrizing contributions to decays of the kind $\ell'\rightarrow
3\ell$ can be written as
\begin{align}
\lag_\text{eff}=-2\sum_{L,R} C_{A B}\, (\bar\ell' P_A \ell)(\bar\ell P_B \ell) \,,
\end{align}
The corresponding decay width is
\begin{align}
\Gamma(\ell'\rightarrow 3 \ell)=\frac{m_\ell^5}{3\cdot 2^{12}\pi^3}\left(|C_{LL}|^2+|C_{RR}|^2+2|C_{LR}^2|+2|C_{RL}|^2\right)\,.
\end{align}
Tree-level contributions from flavon exchange are generated from
diagrams like the one shown on the right in Fig.~\ref{fig:MuEG}. The
corresponding Wilson coefficients read
\begin{align}
C_{LL} = C_{RR}^* 
= g_{\ell\ell'}^*g_{\ell\ell}^*\left(\frac{1}{m_a^2}-\frac{1}{m_s^2}\right)
\qquad \text{and} \qquad 
C_{LR} = C_{RL}^* =
g_{\ell\ell'}^*g_{\ell\ell}\left(\frac{1}{m_a^2}+\frac{1}{m_s^2}\right) \; .
\end{align}
In the case of $\mu\rightarrow 3e$ decays the largest
contribution at one loop, as shown on the left of Fig.~\ref{fig:MuEG},
are suppressed by an additional factor
\begin{align}
\frac{\lambda^2m_\tau}{9m_e m_\mu} \approx 0.1\,,
\end{align}
and therefore negligible. For $\tau \rightarrow 3 \ell$ decays, this
suppression is even more pronounced.\bigskip

The most stringent current bounds on flavor violating three-body
decays are~\cite{tau3mu_tau3e,mu3e}
\begin{align}
\br(\tau \rightarrow 3\mu) &< 2.1 \e{-8} \notag \\
\br(\tau \rightarrow 3e) &< 2.7 \e{-8} \notag \\
\br(\mu \rightarrow 3e) &< 1.0 \e{-12} \; .
\end{align}
Mu3e will improve the limit on $\br(\mu \rightarrow 3e)$ by at least
five orders of magnitude~\cite{Kiehn:2015eta}. However, from flavon
exchange we only expect branching ratios around $\br(\mu \rightarrow
3e)=\mathcal{O}(10^{-20})$, $\br(\tau \rightarrow
3e)=\mathcal{O}(10^{-19})$ and $\br(\tau \rightarrow
3e)=\mathcal{O}(10^{-16})$.  Charged lepton decays with multiple
flavor violations such as $\tau \rightarrow \mu e e $ are further
phase-space suppressed.

\section{Future hadron collider measurements}
\label{sec:coll}

\begin{figure}[t]
\includegraphics[width=0.42\textwidth]{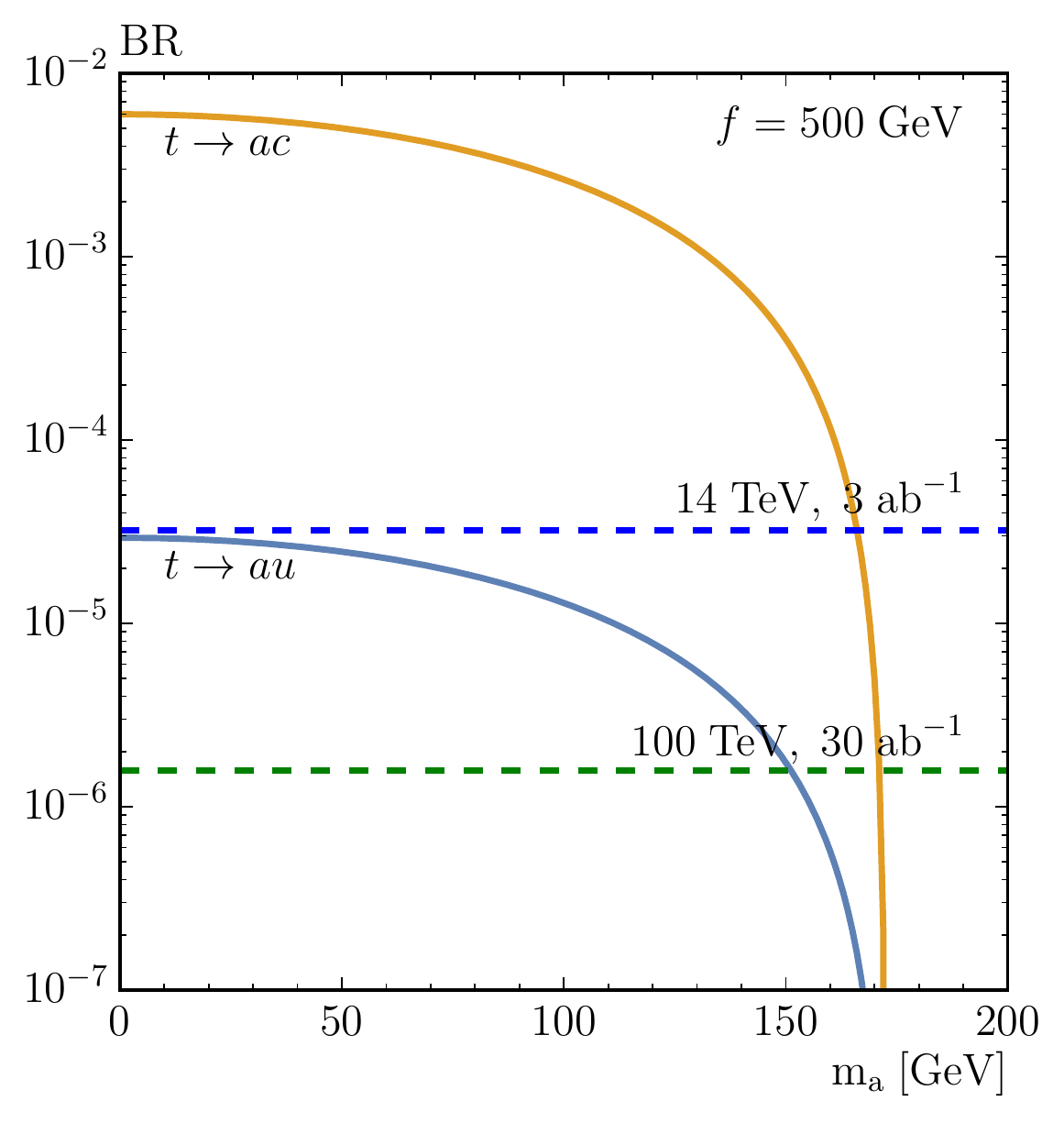} 
\hspace*{0.10\textwidth}
\raisebox{-.1cm}{\includegraphics[width=0.435\textwidth]{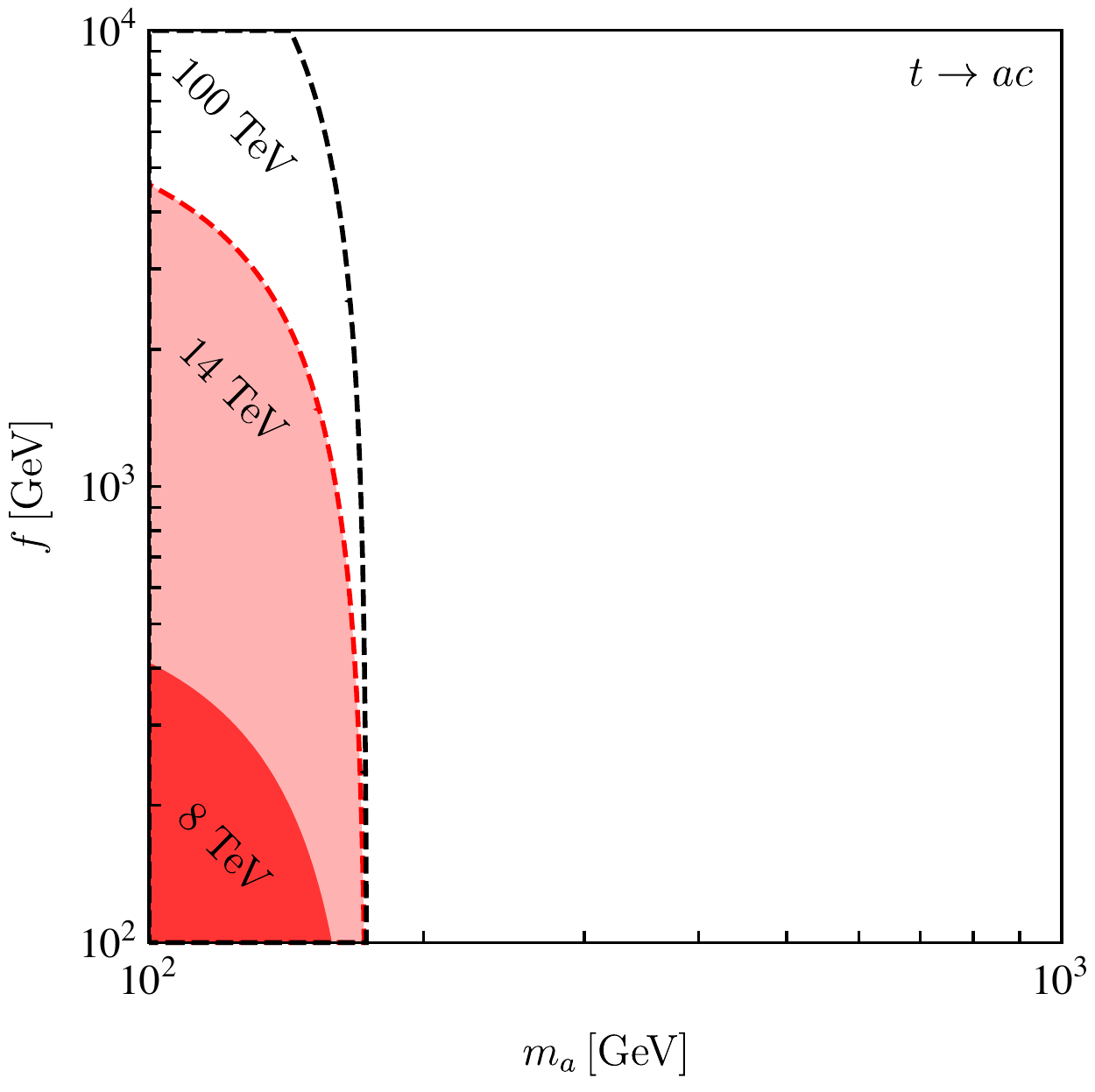}}
\caption{Left: top branching ratios into a flavon and a jet as a
  function of the flavon mass, assuming a fixed VEV of $f =
  500$~GeV. Right: regions in the $m_a-f$ plane excluded by these days
  at the LHC and at a 100~TeV collider.}
\label{fig:topdecay}
\end{figure}

Before we discuss the physics opportunities for flavon searches at a
$100~\tev$ hadron collider, we need to briefly consider limits from
direct LHC searches.  For small flavon masses the main search channel
at hadron colliders are the anomalous top decays given in
Eq.\eqref{eq:topdecay}. The current measurement of the total top width
at the Tevatron gives $1.10~\gev < \Gamma_\text{tot} <
4.05~\gev$~\cite{pdg}. The large error bars indicate that this
global observable will not help searching for flavon
contributions. For the LHC we do not expect this picture to change
significantly. 

Instead, we can search for specific anomalous decays in analogy to the
current limit of $\br(t \to H q) \lesssim 0.5 \%$~\cite{pdg}.  The
current and expected reach for such anomalous top decays at the LHC
and at a 100~TeV hadron collider is~\cite{limit_top_decay}
\begin{align}
\br_{8~\tev}(t\rightarrow Hc)&< 5.6 \e{-3} \notag \\
\br_{14~\tev, 3~\iab}(t\rightarrow Hc)&< 4.5 \e{-5} \notag \\
\br_{100~\tev, 30~\iab}(t\rightarrow Hc)&< 2.2 \e{-6}\,,
\end{align}
based on the channel $H \to b\bar{b}$.  Our estimate for a
$100~\tev$ hadron collider comes from scaling the number of expected tops
by the leading-order ratio of $\sigma (pp \to t \bar{t})$ at $14~\tev$
and $100~\tev$ with \textsc{Madgraph}~\cite{madgraph}.  Assuming a
Gaussian scaling the limit of the counting experiment should improve
by a factor $
6.4 \sqrt{\mathcal{L}_{100~\tev}/\mathcal{L}_{14~\tev}}$.

We can translate these limits into flavon contributions of the kind
$\br( t \to ac \to b\bar{b} c)$, using $\br(a\to b \bar{b}) > 80\%$
from Fig.~\ref{fig:br}.  We show the expected flavon limits as a
function of the flavon mass and couplings in Fig.~\ref{fig:topdecay}
and find that for a 100~TeV they extend to couplings
\begin{align}
\sqrt{ g_{ij}^2 + g_{ji}^2} \; \lesssim  \dfrac{1\e{-3}}{1-\dfrac{m_a}{m_t}} \; .
\end{align}
\bigskip

\begin{figure}[t]
\includegraphics[width=0.42\textwidth]{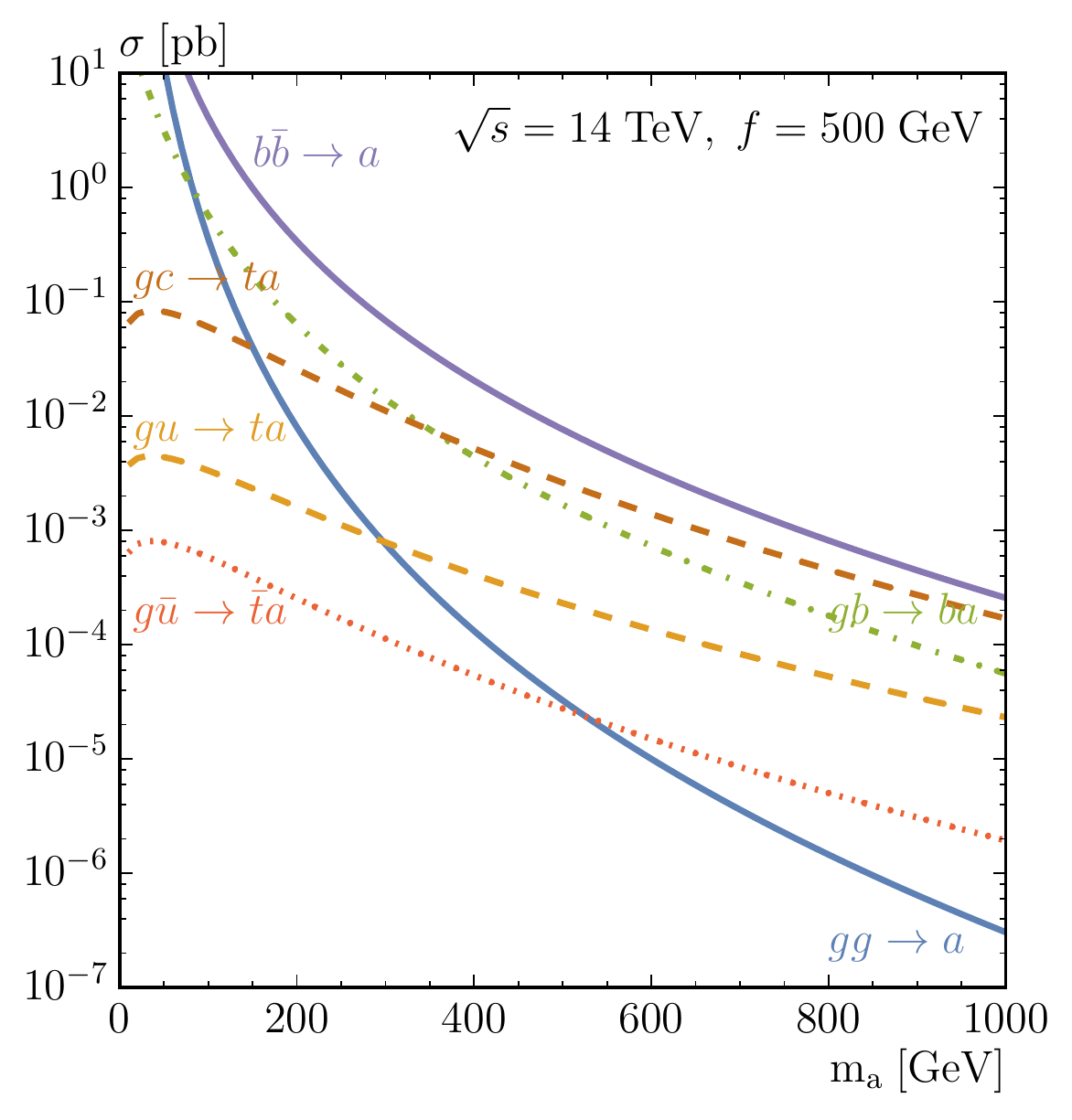}
\hspace*{0.10\textwidth}
\includegraphics[width=0.42\textwidth]{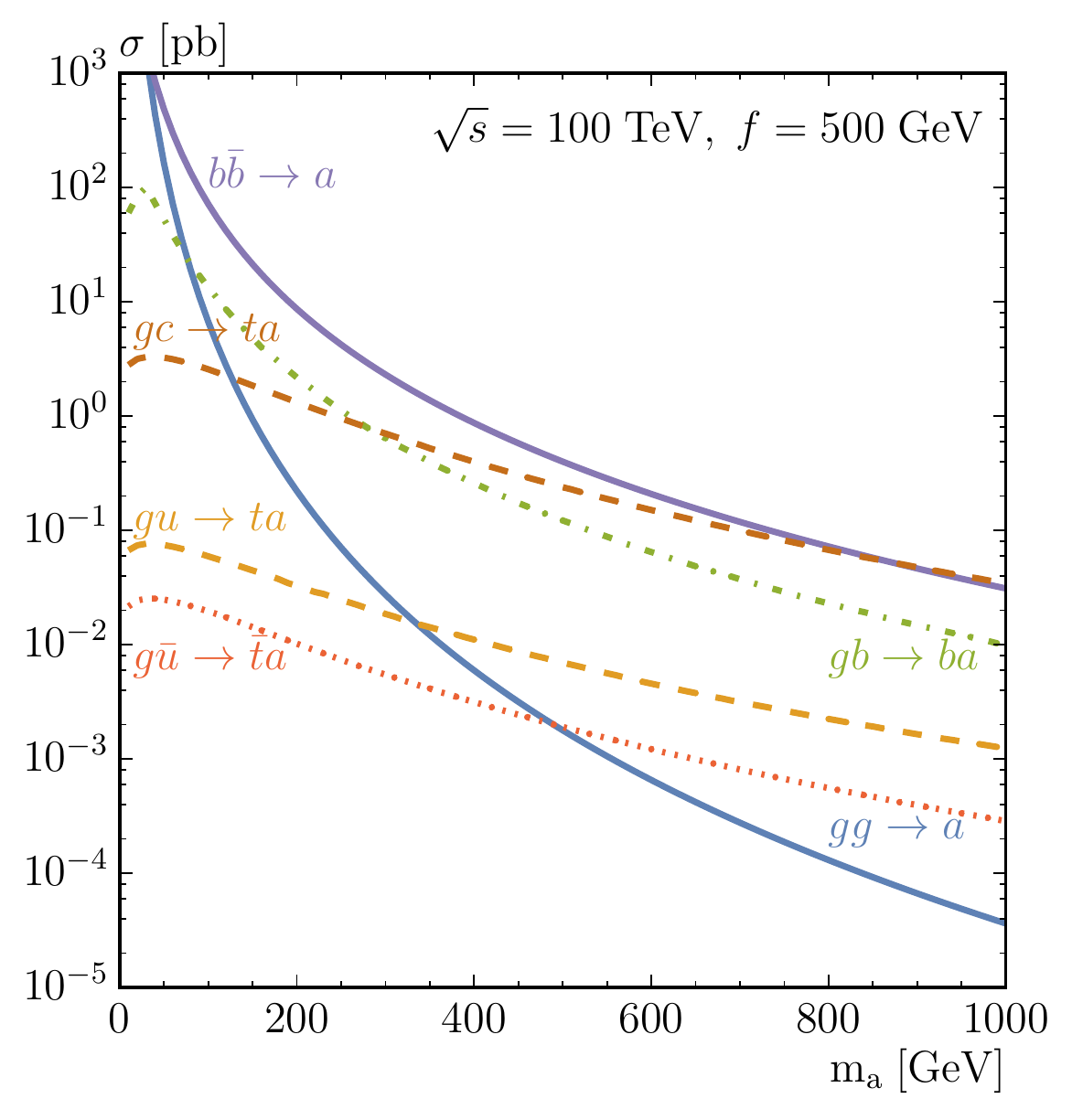}
\caption{Flavon production cross sections in the different channels
  for the $14~\tev$ LHC and a $100~\tev$ hadron collider using the
  \textsc{MSTW2008} PDF set~\cite{mstw}. Couplings are evaluated at
  $\mu = m_a$ or $\mu = m_a+m_t$ with \textsc{CRunDec}~\cite{crundec}.}
\label{fig:prod}
\end{figure}

Next, we compute different flavon production rates at hadron colliders.
Single flavon production occurs as
\begin{align}
gg, b\bar{b} \to a \; ,
\label{eq:prod1}
\end{align}
where we assume that the collinear bottoms in the final state do not
give us an experimental handle on the signal vs background. In
addition, there exist associated production channels
\begin{align}
bg \to ab 
\qqquad \text{or} \qqquad 
ug, cg \to a t \; .
\label{eq:prod2}
\end{align}
Here, we assume the additional $b$-quark to be hard and central, so it
can be tagged. While the bottom-associated channel is driven by a
flavor-diagonal coupling $g_{bb}$, the top-associated production
indicates a flavor-violating flavon-quark coupling. The different
production cross sections for the LHC and for a $100~\tev$ hadron
collider are shown in Fig.~\ref{fig:prod}. At the latter with an
integrated luminosity of $30~\iab$, we would expect to produce
millions of flavons with $m_a > 500~\gev$. This leads us to study two
kinds of collider signatures:
\begin{itemize}
\item first, we can search for traditional resonance decays, like $a
  \to \tau\tau$. In that case all production processes in
  Eqs.\eqref{eq:prod1} and~\eqref{eq:prod2} contribute;
\item second, we can make use of specific top-associated production,
  where the flavon decays into $t\bar{q}$ and $\bar{t}q$ for $q=u,c$ are
  equally likely.
\end{itemize}
In both cases the key question will be how to control large
backgrounds.

\subsubsection*{Resonance searches}

A direct way to search for a flavon as new dynamical degree of freedom
is a resonance search, for example 
\begin{align}
p p \to a \to
b \bar{b} / \tau^+ \tau^- \; .
\end{align}
The $\gamma \gamma$ channel can be discarded, unless we invoke either
a diagonal flavon coupling to the tops or a coupling to the $W$-boson.
To estimate the discovery potential of a $100~\tev$ hadron collider,
we again scale the current 8~TeV LHC limits assuming Gaussian
statistics and an increase of the background cross section by a factor
ten.
%
%
In Tab.~\ref{tab:limits} we show some of the 8~TeV limits together
with our estimate for a 100~TeV hadron collider.  It turns our that
only the flavon channel $p p \to a \to \tau \tau$ may become sensitive
to our benchmark point.\bigskip

\begin{table}[t]
\centering
\ra{1.2}
\begin{tabular}{@{}l|rr|rr|rr|rr@{}}
\toprule
& \multicolumn{2}{c|}{ATLAS 8 TeV} & \multicolumn{2}{c|}{CMS 8 TeV} & \multicolumn{2}{c|}{100 TeV, $30~\iab$} & \multicolumn{2}{c}{benchmark}\\
$m_{a}$~[GeV] & \myalign{c}{500} & \myalign{c|}{1000} & \myalign{c}{500} & \myalign{c|}{1000} & \myalign{c}{500} & \myalign{c|}{1000} &\myalign{c}{500} & \myalign{c}{1000}\\
\midrule
jet-jet~[pb]              &  & $0.2$  &            &           &      &   $2\e{-2}$         & $2.4\e{-2}$ & $1.6\e{-3}$\\
$\tau^+ \tau^-$~[pb]      & $4\e{-2}$ & $5\e{-3}$  & $4\e{-2}$ & $9\e{-3}$ & $3\e{-3}$ & $4\e{-4}$ & $4.1\e{-3}$ & $3.0\e{-4}$  \\
$\mu^+ \mu^-$~[pb]        & $5\e{-3}$ &\phantom{xx}  $1\e{-3}$  & $2\e{-3}$ &\phantom{xx}  $8\e{-4}$ & $2\e{-4}$ &\phantom{xx}  $6\e{-5}$ & $4.0\e{-5}$  &\phantom{xx}  $2.9\e{-6}$ \\
$\gamma \gamma$~[pb]      & $6\e{-3}$ & $1\e{-3}$  & $2\e{-3}$ &           & $2\e{-4}$  & $8\e{-5}$ & $2.3\e{-9}$ & $6.1\e{-11}$\\
\bottomrule
\end{tabular}
\caption{Current~\cite{atlas_gaga,atlas_mumu,atlas_tautau,atlas_jj,cms_jj,cms_gaga,cms_mumu,cms_jj,cms_tautau} 
  and expected limits for $\sigma \times \br$ in pb, assuming an
  increase in the background rate by a factor 10.  For the flavon
  signal we assume $f=500~\gev$.}
\label{tab:limits}
\end{table}

While the above resonances searches are generic for any new (pseudo)scalar,
the off-diagonal flavon coupling $g_{tc,ct}$ introduces a single top signature
\begin{align}
pp \to a \to t\bar{c} / t\bar{u} \; .
\end{align}
The $s$-channel resonance topology only benefits from large branching
ratios for heavy flavons, while the $t$-channel topology suffers from
two flavon couplings. The SM background is single top production with
a NLO cross section $73.5~\pb$ at 100~TeV, requiring $|\eta_t| <
2.5$~\cite{single_top_100tev}. For a flavon with a mass of 500~GeV or
1~TeV we expect for $f = 500~\gev$ $\sigma\times\br = 0.37~\pb$ or
$2.9\e{-2}~\pb$, respectively. Even before considering the price to
pay for charm tagging and without taking into account the top pair
background, we note that this channel will obviously not be sensitive.

\subsubsection*{Associated production}

In addition to these resonance searches, the large flavor-changing
coupling $g_{tc,ct}$ allows for top-associated production,
Eq.\eqref{eq:prod2}. With the relevant flavon decays the collider
signatures are
\begin{align}
pp \to t a \to t \; b\bar{b}/ t \; \tau^+\tau^- \; .
\label{eq:proc_resonance}
\end{align}
The distinctive case of same-sign top production from the decay $a \to
t\bar{c}$ we will treat below.  The decay into bottom quarks suffers
from large combinatorial backgrounds and will be overwhelmed by the
$t\bar{t}$ background. 

\begin{figure}[t]
\includegraphics[width=0.42\textwidth]{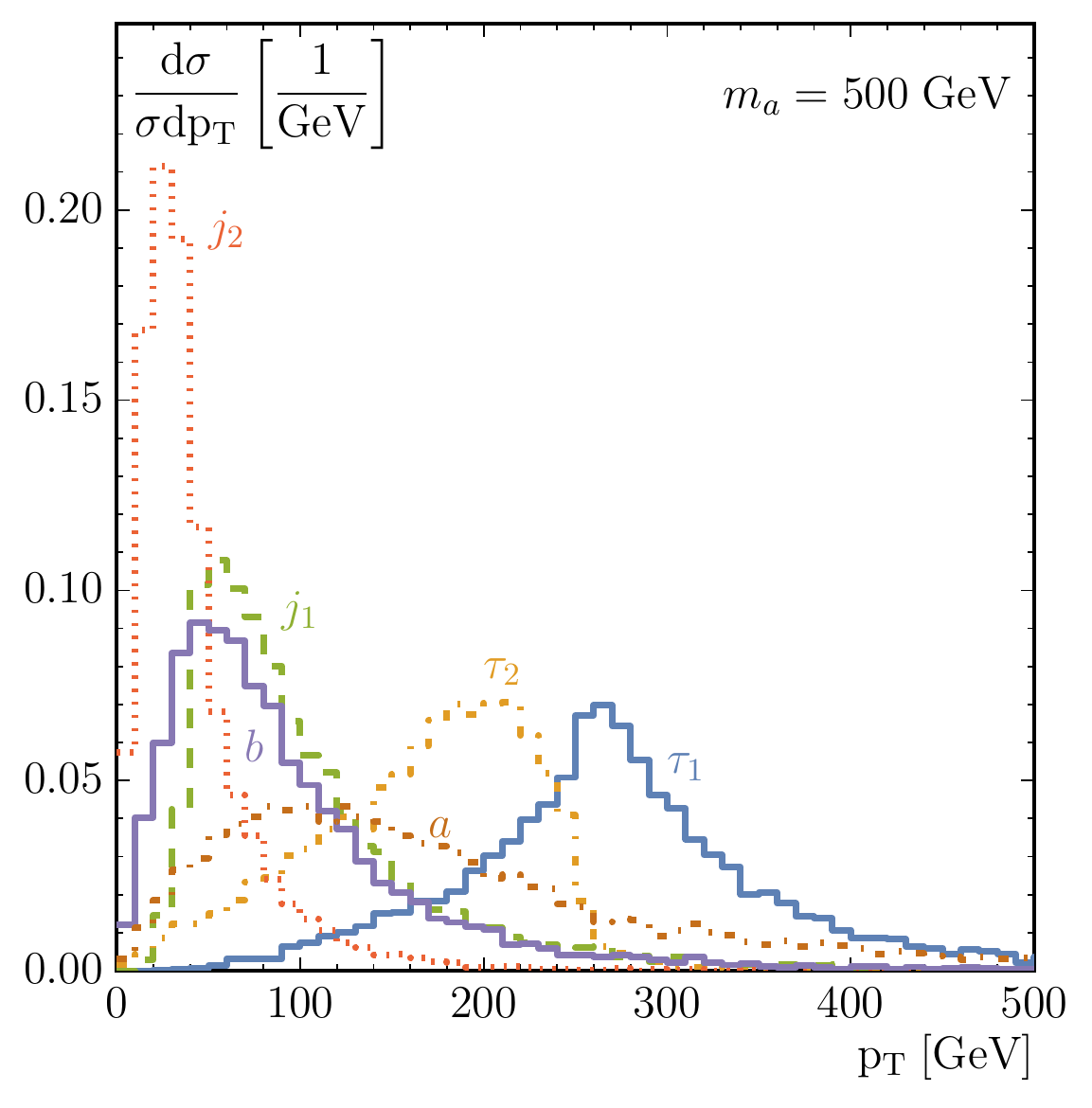}
\hspace*{0.10\textwidth}
\includegraphics[width=0.42\textwidth]{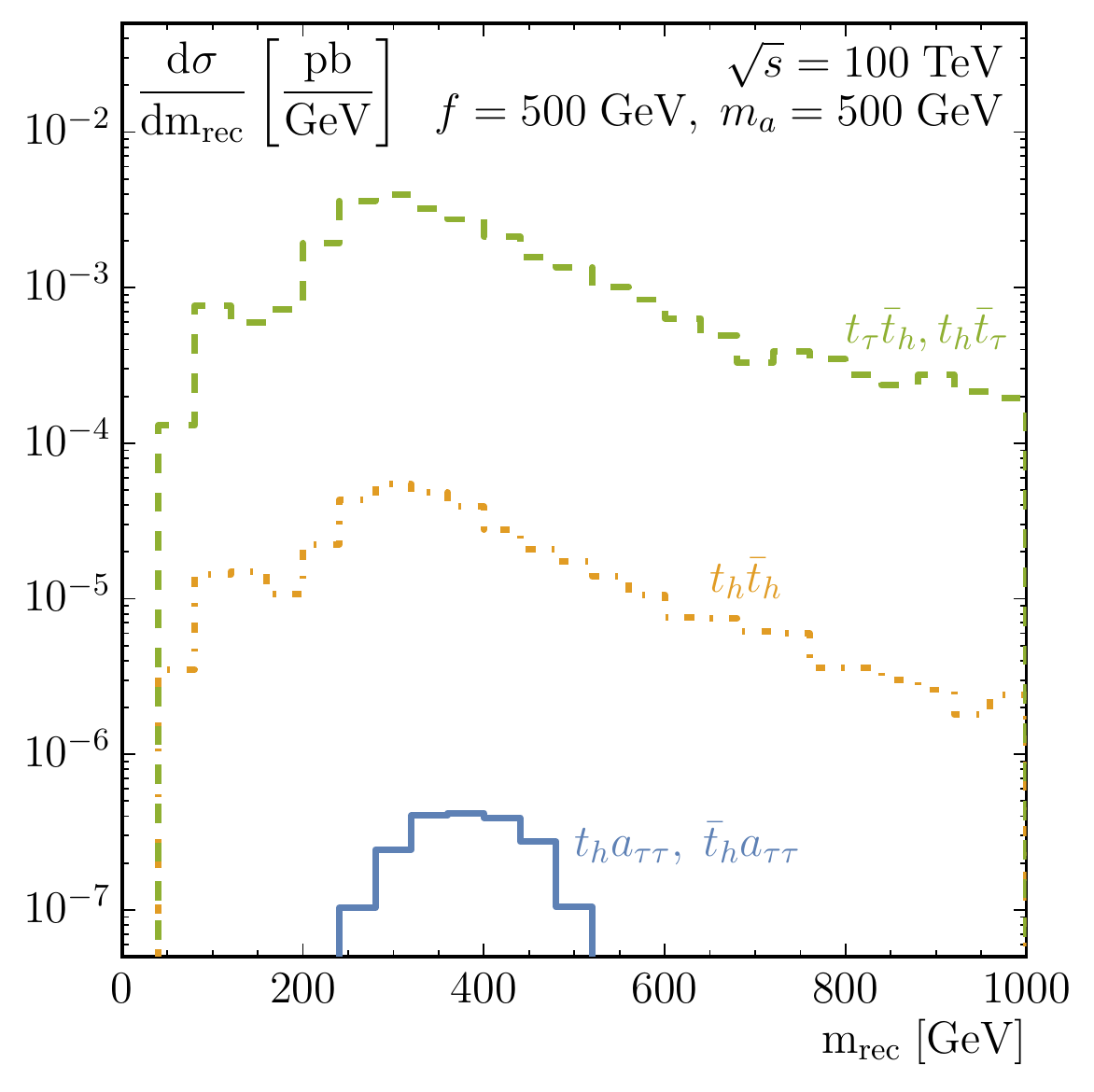}
\caption{Left: parton-level $p_T$ distributions. Right: reconstructed
  mass distribution. Both figures are simulated for $m_a =
  500~\gev$.}
\label{fig:toptautau}
\end{figure}

A flavon decay to (hadronic) taus can be combined with hadronic top
decays, allowing us to reconstruct the final state.  The heavy flavon
would then decay to two boosted taus, significantly harder than the
three top decay jets illustrated in Fig.~\ref{fig:toptautau}. We start
by asking for at least five jets and no isolated leptons,
\begin{align}
  n_j \geq 5 \qqquad 
  n_\ell = 0  \qqquad 
  p_{T,j_1} > 150~\gev \qqquad
  m_{j_3j_4j_5} \in [140,190]~\gev \; .
\end{align}
We assume an optimistic $\tau$-tagging efficiency $\eps_\tau=0.3$ and
a misidentification rate of $\eps_j=10^{-3}$~\cite{atlas_tautagging}.
To reconstruct the flavon we rely on the collinear approximation in
terms of the momentum fractions $x_{1,2}$ of the decaying taus,
\begin{align}
m_{\tau\tau}^2 = \frac{2(p_{j_1}p_{j_2})}{x_1x_2} \; .
\end{align} 
We simulate the flavon signal implemented via
\textsc{FeynRules}~\cite{feynrules} as well as a fully-hadronic
$t\bar{t}$ sample and a $t\bar{t}$ sample with one hadronic top and
the other top decaying to a $\tau$-lepton with
\textsc{MadGraph5}+\textsc{Pythia8} +\textsc{Delphes3}~\cite{madgraph,pythia8,delphes},
employing $R=0.4$ anti-$k_T$ jets from
\textsc{FastJet3}~\cite{fastjet}. For the jets we require
$p_T > 20~\gev$ and $|\eta| < 2.5$. The reconstructed flavon mass
distribution for $m_a = 500~\gev$ and $f = 500~\gev$ is shown in
Fig.~\ref{fig:toptautau}. It is shifted towards lower masses caused by
losses in the reconstruction. The comparison of the expected signal
with the background kills any motivation to further study this
signature.

\subsubsection*{Same-sign top pairs}

As alluded to in Eq.\eqref{eq:proc_resonance}, the most interesting
flavon signature is same-sign top production with an additional jet,
\begin{align}
  pp  \to t_\ell a \to t_\ell t_\ell \bar{c} \; ,
\end{align} 
with a partonic $gc$ initial state.  It leads to two same-sign
leptons, two $b$-jets, and one additional jet. The SM background is $p
p \to b b W^+ W^+ j$, with a leading order cross section of $5.7\cdot
10^{-7}~\pb$. This means that the irreducible background is actually
negligible.  Instead, we need to consider consider $t_{\ell}\bar{t}Z j$ and
$t_{\ell}\bar{t}W^+j$ production, with at least one leptonic top decay and a
leptonically decaying weak bosons.  We simulate the hard process with
\textsc{Madgraph5}+\textsc{Pythia8}+\textsc{Delphes3}~\cite{madgraph,pythia8,delphes,detector_card}. The
expected flavon signal has a rate of $5.4\e{-3}~\pb \times
(500~\gev/f)^2$ for $m_a =500~\gev$. The two leading background are
significantly larger, $\sigma_{t_{\ell}\bar{t}W^+j} = 0.33~\pb$ and $\sigma_{t_{\ell}\bar{t}Zj} =
0.48~\pb$.\bigskip

\begin{figure}[t]
\centering
\includegraphics[width=0.42\textwidth]{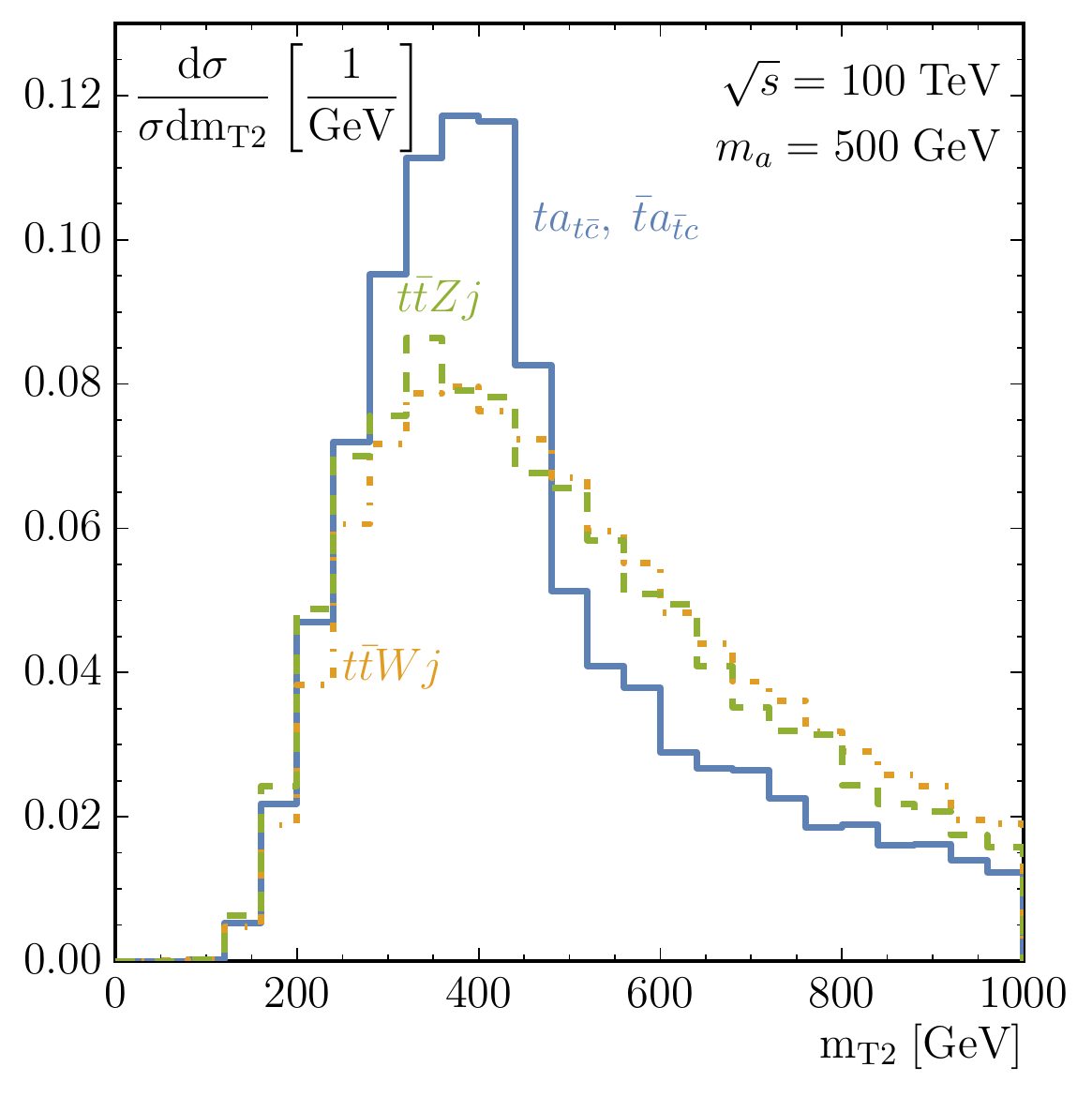}
\hspace*{0.10\textwidth}
\raisebox{-.1cm}{\includegraphics[width=0.435\textwidth]{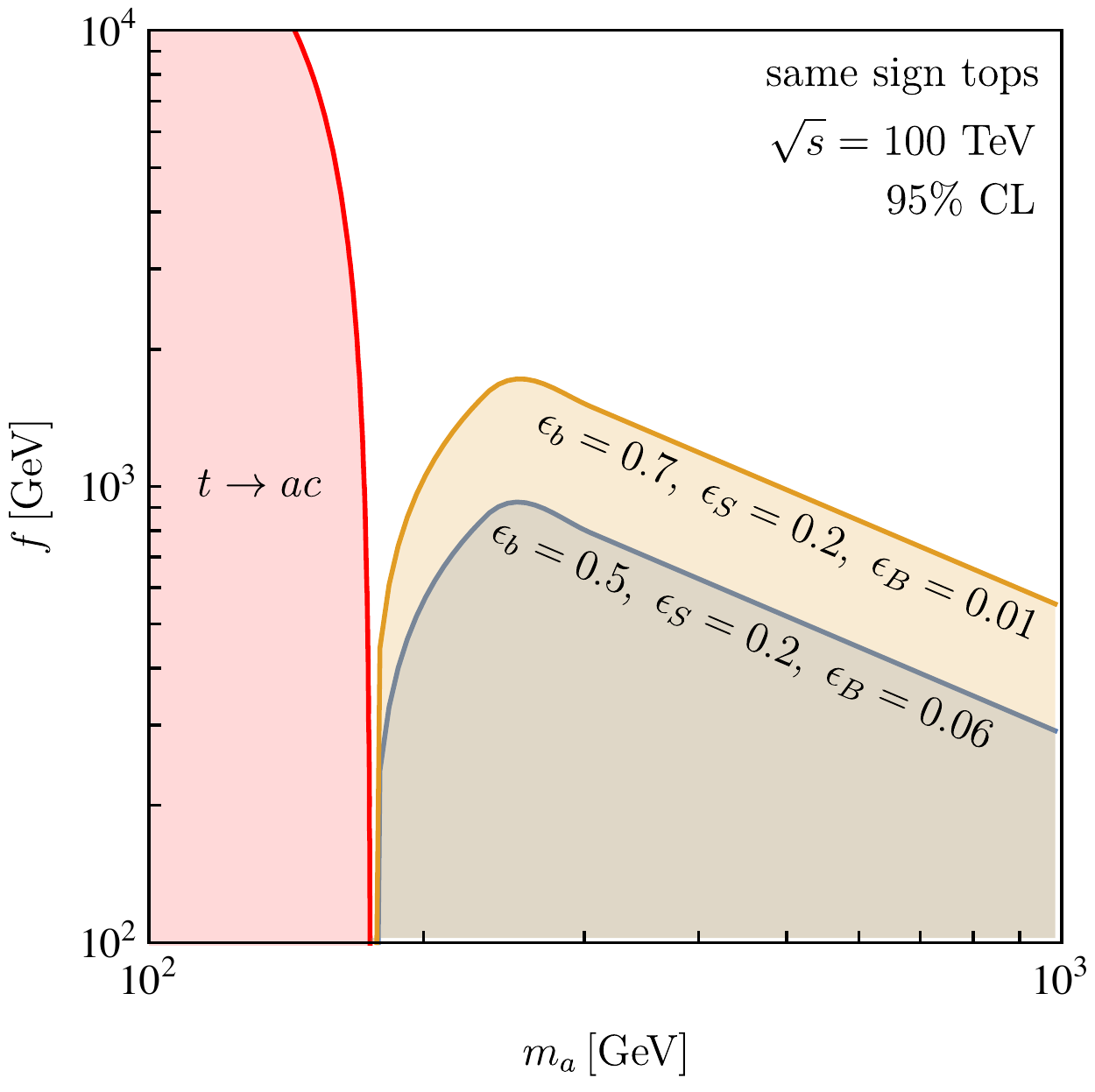}}
\caption{Left: normalized $m_{T2}$ distribution for a $m_a = 500~\gev$
  flavon and backgrounds. Right: Exclusion limits from
  $\sigma(g c \to t a) \times \br(a\to t \bar{c})$[pb] at
  $\sqrt{s} = 100~\tev$. The red area is excluded by $t \to a c$
  decays.}
\label{fig:sst_gc}
\end{figure}

We require two isolated same-sign leptons with
\begin{align}
R_\text{iso} = 0.2 \qqquad 
I_\text{iso} =0.1 \qqquad 
p_{T,\ell} > 10~\gev \qqquad 
|\eta_\ell| < 2.5 \; . 
\end{align}
In events with more than two such leptons we pick the hardest two. We
veto events with a third lepton of different sign and one
opposite-sign combination fulfilling $|m_{\ell^+ \ell^-} - m_Z|<
15~\gev$ to reduce the $t_{\ell}\bar{t}Zj$ background.  The hadronic activity
is clustered into $R=0.4$ anti-$k_T$ jets with $p_T > 40~\gev$ and
$|\eta_j| < 2.5$ using \textsc{FastJet3}~\cite{fastjet}.  The hardest
jet with $p_{T,j} > 100~\gev$ is our $c$-candidate. Among the non-$c$
jets we require at least two $b$-tags with a parton-level $b$-quark
within $R < 0.3$ and an assumed tagging efficiency 50~\%.  Finally, we
target the two neutrinos by requiring $\slashed{p}_T > 50~\gev$. This
missing transverse momentum has to be distributed between the two
branches of the event, the flavon decay and the top decay. A powerful
observable for such topologies is $m_{T2}$~\cite{mt2}.  We define two
branches by assigning each $b$-quark to the leptons and minimizing
$\Delta R_{\ell_1b_i}+\Delta R_{\ell_2b_j}$.  Then we assign the hard
$c$-jet to the top candidate with the smaller $\Delta y_{(\ell
  b),j}$. For most signal events we expect $m_t < m_{T2} < m_a$, which
allows us to search for an excess of events over the background that
provides side-bands at high value of $m_{T2}$. We show the
corresponding distribution in the left panel of Fig.~\ref{fig:sst_gc}.

A final, distinctive feature of the signal is that both leptons
originate form tops, so the two $b$-jets should be tagged with the
same charge~\cite{b_charge}. Recent ATLAS
studies~\cite{b_tagging_charge} show that a $b$-$\bar{b}$ distinction
is possible with $\epsilon_S = 0.2$ and $\epsilon_B = 0.06$. For our
analysis we assume two scenarios: for a conservative estimate we use
these ATLAS efficiencies; for a more optimistic case we assume an
improved mis-tagging rate of $\epsilon_B = 0.01$ and an overall
$b$-tagging efficiency of 70~\%. The obtained exclusion limits at
95~\% CL with the additional requirement $S/B > 0.1$ are illustrated in the
right panel of Fig.~\ref{fig:sst_gc}.

\subsubsection*{Flavon pair production}

In principle, there is a possibility to study flavon pair production.
For the dominant production mode $b \bar{b} \to ff$ with $g_{aabb} = 2
m_b/f^2 $ and $m_a = f = 500$~GeV we find a production cross section
of $1.4\e{-3}$~pb at 100~TeV. The four-$b$ final state will be
overwhelmed by combinatorics and QCD backgrounds. The same-sign $tc \,
tc$ channel offers a more distinct signature, but is overwhelmed by a
$t\bar{t}W^+jj$ cross section of $4.6$~pb, where we require two jets
with $p_{T,j}>100$~GeV. In addition, this channel would not allow for
the reconstruction of a mass peak. Therefore, also flavon pair
production can unfortunately be removed from the list of promising
discovery channels at a 100~TeV hadron collider.

\section{Outlook}
\label{sec:conc}

The experimental consequences of flavor physics models including
flavons have been appropriately described in terms of an effective
field theory for many decades. In the coming decades we will have the
opportunity to test flavon models with a mixed approach of indirect
and direct searches, which has been extremely successful in the case
of the weak gauge bosons, the top quark, and most recently the Higgs
boson.\bigskip

\begin{figure}[t]
\begin{center}
  \includegraphics[width=.42\textwidth]{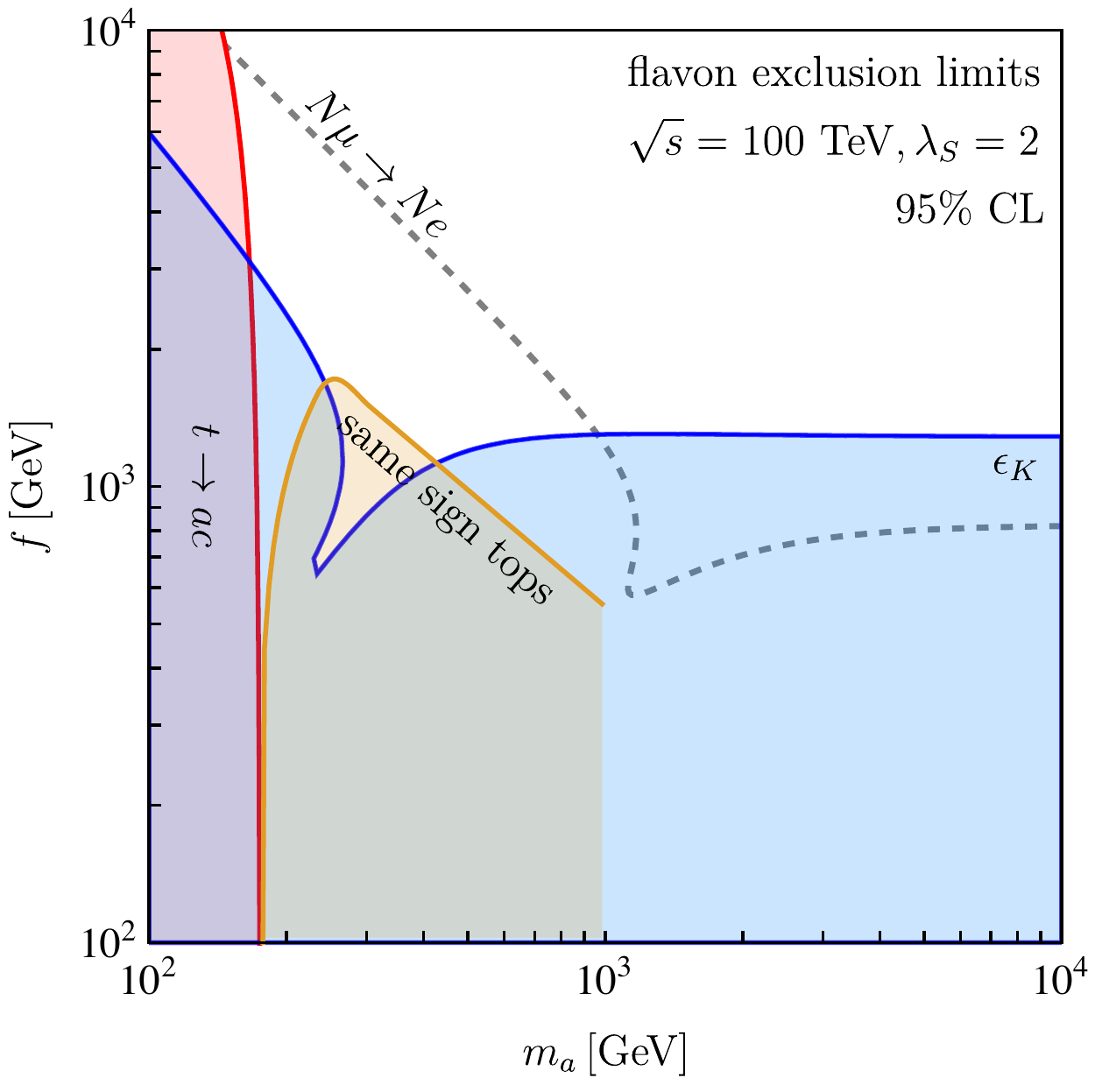}
\end{center}
  \caption{Regions in the $m_a-f$ plane which can be probed by quark
    flavor physics ($\epsilon_K$), by lepton flavor physics ($\mu \to
    e$ conversion), and by a 100~TeV hadron collider. For the latter
    we show the reach of anomalous top decays and same-sign top
    production.}
  \label{fig:outlook}
\end{figure}

Starting with current and future quark flavor physics constraints we
have shown that a large region in the flavon parameter space is
waiting to be probed by alternative experimental approaches in
particle physics. While the projected improvements in the quark flavor
sector, based on meson mixing and rare decays, are of order-one,
lepton flavor experiments should realize their huge potential in the
coming years. In Fig.~\ref{fig:outlook} we show how based on our
benchmark point the indirect searches for lepton flavor effects will
gain immense sensitivity.\bigskip

In addition to indirect searches in the quark and lepton sectors,
systematic direct searches for flavons will for the first time be
possible at a 100~TeV hadron collider. Two kinds of collider
signatures appear for flavons: first, all colliders search for generic
resonant (pseudo-)scalar states, \textit{i.e.} powerful signatures
without a flavon-specific flavor structure. Second, same-sign top pair
production with an additional jet coming from the flavon decay $a \to
tj$, combining a distinctive signature with a slightly more
background-prone resonance structure. Tools like bottom vs anti-bottom
tagging would be extremely useful to extract such signatures at future
colliders. In Fig.~\ref{fig:outlook} we show how the projections for a
100~TeV collider nicely add to the indirect searches. Both, quark
flavor and lepton flavor searches show distinctive dips close to the
diagonal $m_a \sim f$, driven by a destructive interference of virtual
scalar and pseudo-scalar contributions. On the quark side, anomalous top decays have an
excellent coverage for small flavon masses, while the same-sign top
channel can cover exactly the weak parts of the indirect searches
around $m_a \sim f$. In combination, the different tests clearly allow
for a systematic and independent coverage of the flavon parameter space in the leptonic sector as well as in the hadronic sector. Ideally this 
includes a direct discovery of flavon-specific couplings at the
100~TeV hadron collider.

\subsubsection*{Acknowledgments}

First of all, we would like to thank J\"org J\"ackel and Jamie
Tattersall for their help in an early phase of the project.  We are
grateful to Jure Zupan and Andi Weiler for useful discussions
regarding flavon pair production. M.B.\ acknowledges the support of
the Alexander von Humboldt Foundation. T.S.\ would like to thank the
International Max Planck Research School for \textsl{Precision Tests
  of Fundamental Symmetries} for their support.

\newpage 
\appendix
\section{Benchmark Point}

To find sample parameter points, we generate random fundamental Yukawa
couplings with $y_{ij}^{u,d}= |y_{ij}^{u,d}| \,e^{i\phi^{u,d}_{ij}}$
and $|y_{ij}^{u,d}|\in [0.5,1.5]$ and $\phi^{u,d}_{ij}\in [0,2
  \pi]$. The effective Yukawa couplings defined in Eq.\eqref{eq:lag2}
have to reproduce the quark and lepton masses, and mixing angles at
the flavor breaking scale, which we take to be $1$ TeV. For the
numerical values we use
Refs.~\cite{Xing:2007fb,GonzalezGarcia:2012sz}. To this end we perform
a $\chi^2$ fit, with symmetrized $2 \sigma$ errors and require $\chi^2
< 1/\text{d.o.f.}$. To illustrate the results in this paper we define
a benchmark point with the masses
\begin{align}
m_{u_i}&=\left(0.00138, 0.563, 150.1\right)~\text{GeV}\notag \\
m_{d_i}&=\left(0.00342, 0.054, 2.29\right)~\text{GeV}\notag \\
m_{\ell_i}&=\left(0.000513, 0.106, 1.81\right)~\text{GeV}\notag \\
m_{\nu_i}&=\left(0.00161, 0.523, 3.79\right)\e{-11}~\text{GeV}\,,
\end{align}
and the mixing matrices
\begin{align}
|V_\text{CKM}|=\begin{pmatrix}0.974&0.226&0.0035\\
0.226&0.974&0.0388\\
0.011&0.037&0.999
\end{pmatrix}\,,\qquad |V_\text{PMNS}|=\begin{pmatrix}0.813&0.565&0.142\\
0.483&0.519&0.705\\
0.324&0.642&0.695
\end{pmatrix}\,.
\end{align}
The corresponding Yukawa couplings in the quark sector are
\begin{align}
y_u&=\begin{pmatrix}
0.34+0.82i&-0.23+0.69i&0.41-0.43i\\
-0.84+0.26i&-0.64+0.32i&1.35-0.24i\\
0.98-0.90i&-0.84-1.20i&0.75+0.65i
\end{pmatrix}\notag \\
y_d&=\begin{pmatrix}
0.53+0.72i&0.50-0.34i&0.65-0.10i\\
1.12-0.14i&0.93-0.54i&-0.31-0.65i\\
-0.16+0.6i&-0.73+0.34i&0.84+0.61i
\end{pmatrix}\,,
\end{align}
while the lepton sector is described by 
\begin{align}
y_\nu&=\begin{pmatrix}
-0.73-0.49i&0.91-0.68i&0.50-0.21i\\
0.77+0.36i&0.59+0.84i&0.23-1.19i\\
-0.29+1.14i&-0.02-0.59i&1.15+0.91i
\end{pmatrix} \notag  \\
y_\ell&=\begin{pmatrix}
0.16+1.29i&-0.95-0.97i&0.25+0.92i\\
0.008-0.99i&1.11+0.40i&0.47+0.48i\\
0.30-1.30i&0.22+0.77i&-0.59-0.018i
\end{pmatrix}\,.
\end{align}
We note that this benchmark point is not optimized to illustrate
specific features linked to quark flavor, lepton flavor, and collider
reaches. The quark flavor and collider sector on the one hand, and the
lepton sector on the other are only loosely related. All couplings are
deliberately chosen in the weakly interacting regime, to avoid
conclusions too closely tied to assumptions about underlying
ultraviolet completions.


\end{document}

%% file: declare.tex
\newcommand{\e}[1]{\cdot 10^{#1}}
\newcommand{\eps}{\epsilon}
\newcommand{\Heff}{{\cal H}_\text{NP}}

\newcommand{\lag}{\ensuremath{\mathcal{L}}}

\newcommand{\br}{\text{BR}}

\newcommand{\qqquad}{\qquad \qquad}

\newcommand{\gev}{{\ensuremath\rm GeV}}
\newcommand{\tev}{{\ensuremath\rm TeV}}

\newcommand{\pb}{{\ensuremath\rm pb}}

\newcommand{\iab}{{\ensuremath\rm ab^{-1}}}

\def\slashchar#1{\setbox0=\hbox{$#1$}           
   \dimen0=\wd0                                 
   \setbox1=\hbox{/} \dimen1=\wd1               
   \ifdim\dimen0>\dimen1                        
      \rlap{\hbox to \dimen0{\hfil/\hfil}}      
      #1                                        
   \else                                        
      \rlap{\hbox to \dimen1{\hfil$#1$\hfil}}   
      /                                         
   \fi}
